\documentstyle[aps,prb,twocolumn,epsf,floats]{revtex}
\begin{document}
\preprint{DRAFT}

\twocolumn[\hsize\textwidth\columnwidth\hsize\csname @twocolumnfalse\endcsname

\draft

\title{Local fluctuations in quantum critical metals}

\author{Qimiao Si$^\ast$, Silvio Rabello$^\ast$,
Kevin Ingersent$^\dag$, and J.\ Lleweilun Smith$^\ast$
}
\address{$^\ast$Department of Physics \& Astronomy, Rice University, Houston,
TX 77251--1892, U.S.A. \\
$^\dag$Department of Physics, University of Florida, Gainesville,
FL 32611--8440, U.S.A.}

\maketitle
\begin{abstract}

We show that spatially local, yet low-energy, fluctuations can play an
essential role in the physics of strongly correlated electron systems
tuned to a quantum critical point.
A detailed microscopic analysis of the Kondo lattice
model is carried out within an extended dynamical mean-field
approach. The correlation functions for the lattice model are
calculated through a self-consistent Bose-Fermi Kondo problem,
in which a local moment is coupled both to a fermionic bath and
to a bosonic bath (a fluctuating magnetic field).
A renormalization-group treatment of this impurity problem---perturbative in
$\epsilon=1-\gamma$, where $\gamma$ is an exponent characterizing the
spectrum of the bosonic bath---shows that competition between the
two couplings can drive the local-moment fluctuations critical.
As a result, two distinct types of quantum critical point emerge in the Kondo
lattice, 
one being of the usual spin-density-wave type, the other ``locally critical.''
Near the locally critical point, the dynamical spin susceptibility exhibits
$\omega/T$ scaling with a fractional exponent.
While the spin-density-wave critical point is Gaussian, the
locally critical point is an interacting fixed point
at which long-wavelength and spatially local critical modes coexist.
A Ginzburg-Landau description for the locally critical point is
discussed.
It is argued that these results are robust, that local criticality provides
a natural description of the quantum critical behavior seen in a number of
heavy-fermion metals, and that this picture may also be relevant to other
strongly correlated metals.

\end{abstract}
]

\section{Introduction}
\label{sec:intro}

Non-Fermi-liquid properties have been seen experimentally in
a number of strongly correlated electron systems.%
\cite{StewartRMP,Timusk,VarmaNFL}
They pose fundamental questions about how electron correlations
lead to new electronic states of matter and new elementary
excitations. One mechanism for the breakdown of Fermi-liquid theory
is quantum criticality. 
While the extent to which an underlying quantum critical point (QCP)
plays a role in the non-Fermi-liquid behavior of high-temperature
superconductors remains a subject of debate,\cite{Tallon}
the situation is much clearer in heavy-fermion metals. 
Among the many heavy-fermion materials in which non-Fermi-liquid
properties have been seen,\cite{StewartRMP,vonLohneysen,Lonzarich,Steglich,%
Stewart,Flouquet,Thompson,deVisser,Maple,Andraka,Aronson,MacLaughlin}
magnetic QCPs have been explicitly
identified in a number of stoichiometric or nearly stoichiometric
materials\cite{vonLohneysen,Lonzarich,Steglich,Stewart} including
CeCu$_{6-x}$Au$_x$, CePd$_2$Si$_2$, CeIn$_3$, and YbRh$_2$Si$_2$.
In each of these systems, the
zero-temperature
magnetic ordering transition appears to be continuous.
Non-Fermi-liquid behavior, usually seen in the temperature dependences
of transport and thermodynamic properties, arises near the QCP.
For instance,
the resistivity is linear (or close to being linear)
in $T$, and the temperature dependence of the specific heat coefficient
$C/T$
is either logarithmically singular or nonanalytic with 
a finite zero-temperature limit.
Away from the quantum critical regime, there is a gradual recovery
of Fermi-liquid behavior (albeit still with a large effective mass).
These materials provide controlled settings in which to study not only
basic issues concerning magnetism in heavy fermions,
but also the physics of QCPs in strongly correlated metals in general.

Inelastic neutron scattering
directly probes
the critical
fluctuations at heavy-fermion magnetic QCPs.
Particularly detailed measurements have been performed on
CeCu$_{6-x}$Au$_x$, which can be tuned from a paramagnetic metal
to an antiferromagnetic metal by varying $x$
and/or an applied pressure.
The results\cite{Schroder2,Schroder1,Stockert,Stockert2}
are striking
in a number of ways:
(1)~The frequency dependence of the dynamical
spin susceptibility displays a fractional exponent.
(2)~The same exponent also appears in the temperature dependence;
the dynamical spin susceptibility has $\omega/T$ scaling.
(3)~This
fractional exponent describes the variation with frequency and
temperature not only at the ordering wavevector,
but essentially everywhere else in the Brillouin zone.

The existence of the fractional exponent
is a surprise.
It is standard to 
assume\cite{Hertz,Millis,Lonzarich-review,Moriya,Continentino,Lavagna}
that the critical theory for
magnetic QCPs in metals is a $\phi^4$ theory, describing
the long-wavelength fluctuations of the magnetic order parameter.
Landau damping makes the dynamic exponent $z$ larger than 1.
In the antiferromagnetic case, $z=2$, so the effective dimensionality of the
$\phi^4$ theory, $d_{\text{eff}}=d+z$, is greater than (equal to) its upper
critical dimension for spatial dimensionality $d=3$ ($d=2$).
Consequently,
in the standard picture
all the nonlinear couplings are irrelevant in the
renormalization-group (RG) sense, the fixed point is Gaussian,
and the frequency exponent must take its mean-field value of 1.

Likewise, the existence of $\omega/T$ scaling is surprising.
For a Gaussian fixed point,
the spin relaxation rate is determined by nonlinear couplings that
are irrelevant in the RG sense.
This implies that the relaxation rate is superlinear
in temperature,\cite{Sachdev-book} whereas $\omega/T$ scaling
can arise only for a linear relaxation rate.

Both of these features, then, imply that the critical point in
CeCu$_{6-x}$Au$_x$ has to be an interacting one.
What is the new physics that is responsible
for the interacting part of the critical theory? A clue is provided by
the third feature of the experimental data mentioned
above.
The fact that the fractional exponent in the frequency/temperature dependence
occurs also at generic wavevectors (far away from the ordering
wavevector) suggests that the origin of the fractional exponent lies
in some type of local
physics.\cite{Schroder1}

Heavy-fermion metals undergoing a magnetic quantum phase transition should
be well-described by the Kondo lattice model of local moments interacting
with conduction-band electrons.
It is natural, therefore, to suspect that the local physics responsible for
the anomalous properties of CeCu$_{6-x}$Au$_x$
involves the magnetic moments.
In Ref.~\onlinecite{Nature}, we briefly reported
an interacting critical point in Kondo lattice systems.
(See also Ref.~\onlinecite{SiSmithIngersent}
for more qualitative discussions.)
Our analysis was carried out using an extended dynamical mean field
theory (EDMFT), supplemented by Ginzburg-Landau considerations.
Other approaches, based on a large-$N$ formulation, have also
been developed to study this problem,\cite{Coleman1,Coleman2}
although they have yet to yield any new type of QCPs.
A scenario
of three-dimensional electrons
coupled to two-dimensional Gaussian spin
fluctuations\cite{Stockert2,Rosch,Lonzarich}
has also been proposed, but it does not address
the local and other non-Gaussian aspects of the critical
dynamics.

The present paper describes in detail our microscopic analysis
of the Kondo lattice.
The interplay of local and spatially extended physics is treated using
the EDMFT developed in Refs.~\onlinecite{Smith1,Smith2,Chitra}.
The central result is the finding of two types of quantum phase transition.
The first, corresponding to the Gaussian picture, describes a transition
of the usual spin-density-wave (SDW) type,
and is found to occur when the underlying spin fluctuations are
three-dimensional in character.
We dub the other type of transition ``locally critical,''
because long-wavelength and spatially local critical modes coexist at
the QCP.
This type of transition occurs if the spin fluctuations are
(quasi-)two-dimensional.
Near the locally critical point, the dynamical spin susceptibility exhibits
$\omega/T$ scaling with a fractional exponent.
By analyzing the general form of the interacting critical theory, we
argue that the existence of two types of critical point is valid beyond
the approximations contained in
our
microscopic calculations.
The locally critical picture provides a natural description of the quantum
critical behavior seen in a number of heavy-fermion metals,
and is argued to be of importance in the broader context of strongly
correlated electron physics.

The remainder of the paper is organized as follows.
We introduce the model and
its EDMFT formulation 
in Section~\ref{sec:edmft-kondo}. 
The EDMFT equations have the content of a
Bose-Fermi Kondo model supplemented by self-consistency conditions.
In Section~\ref{sec:impurity}, we present a detailed RG analysis 
of the Bose-Fermi Kondo model, when the bath spectral functions are
assumed to take power-law forms.
A critical point is identified in the impurity model, and the
spin correlation functions near this critical point are calculated.
Using this solution to the impurity model, we study in 
Section~\ref{sec:self-consistent} the self-consistent
EDMFT problem.
Two types of
QCP
are identified, and their zero-temperature dynamics are determined.
The finite-temperature
spin dynamics are studied in Section~\ref{sec:finite-T-lcp}.
A Ginzburg-Landau analysis is presented in Section~\ref{sec:GL}.
Section~\ref{sec:expt} compares our theoretical results with
experiments in heavy-fermion systems.
Section~\ref{sec:conclusion} contains concluding remarks
and discusses
the general relevance of critical local physics.
Details of the derivation of the RG equations, of the determination of
the separatrix between the two stable phases of the Bose-Fermi Kondo model,
and of the calculation of correlation functions at the critical point of the
impurity model are given in Appendices~\ref{sec:appen-rg},
\ref{sec:appen-separatrix}, and \ref{sec:appen-corr}, respectively.
Finally, Appendix~\ref{sec:appen-spherical} presents the derivation of
correlation functions for the long-ranged one-dimensional
spherical model with an interaction that decays with distance 
either as a pure power-law or with logarithmic corrections.

\section{The model and formalism}
\label{sec:edmft-kondo}

\subsection{Kondo Lattice Model}

The Kondo lattice model is specified by the Hamiltonian
\begin{eqnarray}
{\cal H}
= \sum_{ ij\sigma} t_{ij} ~c_{i\sigma}^{\dagger} c_{j\sigma}
+ \sum_i J_K ~{\bf S}_{i} \cdot {\bf s}_{c,i}
+ \sum_{ ij} { I_{ij} \over 2} 
~{\bf S}_{i} \cdot {\bf S}_{j} .
\label{kondo-lattice}
\end{eqnarray}
At each lattice site $i$ is located a spin-${1 \over 2}$
local moment ${\bf S}_{i}$, which interacts
on-site with the spin ${\bf s}_{c,i}$
of the conduction ($c$) electrons.
The tight-binding parameters $t_{ij}$ determine the dispersion
$\epsilon_{\bf k}$ and, hence, the bare conduction-band density of states
\begin{eqnarray}
\rho_0 (\epsilon)= \sum_{\bf k} \delta ( \epsilon - \epsilon_{\bf k} ) .
\label{bare-cond-electron-dos}
\end{eqnarray}
We will
treat
only cases where the average number of conduction
electrons is less than one per site. All the phases described
below are metallic.

Two processes oppose each other in the Kondo lattice.
First, each magnetic moment couples locally with strength $J_K$
to the spins of the conduction electrons.
We consider only positive (antiferromagnetic) $J_K$ values,
which can lead to Kondo-quenching of the local moments.
Second, the magnetic moments interact with each other through the nonlocal,
RKKY coupling $I_{ij}$.
In Eq.~(\ref{kondo-lattice}), we have introduced an
explicit RKKY interaction term
in addition to that induced implicitly by the Kondo coupling $J_K$.
This is done so that the dynamics associated with the RKKY
interactions can be incorporated into the EDMFT.\cite{Smith1,Smith2,Chitra}
(The EDMFT ensures that there is no double-counting of the implicit
and explicit RKKY interactions.\cite{Smith1})
For the systems under consideration,
$I_{\bf q}$, the spatial Fourier transform of $I_{ij}$, is most negative
when ${\bf q}$ is equal to the antiferromagnetic wavevector, ${\bf Q \ne 0}$.

The competition between the Kondo effect and the RKKY interaction
is expected to lead to a magnetic quantum phase transition.\cite{Doniach,Varma}
Traditionally, it is assumed that the local moments are quenched,
not only on the paramagnetic side but also through the transition.
In this picture, the quantum phase transition corresponds to an SDW
instability of heavy quasiparticles produced by the quenching, and
the quantum critical behavior is necessarily that of a Gaussian fixed point.
By contrast, our analysis makes no prior assumption
that Kondo resonances are fully developed.
It should be stressed---as a discussion at the end of
Section~\ref{sec:self-consistent2D}
will make clear---that the critical point is always reached by
tuning just one parameter: the ratio of the effective RKKY interaction
and an effective Kondo coupling.

\subsection{Extended dynamical mean field theory}

In the EDMFT,\cite{Smith1,Smith2,Chitra} all the correlation functions
of the Kondo lattice can be calculated in terms of a self-consistent
Kondo impurity model.
The EDMFT generalizes the standard dynamical mean-field theory\cite{Georges}
in that it incorporates the quantum fluctuations
produced
by any intersite
interaction; in the Kondo lattice model~(\ref{kondo-lattice})
this is the RKKY interaction.
A key quantity introduced in the EDMFT is the ``spin self-energy''
$M(\omega)$, in terms of which the momentum-dependent
dynamical spin susceptibility can be written
\begin{eqnarray}
\chi ({\bf q}, \omega) &=& \frac {1}  { M(\omega) + I_{\bf q} } .
\label{chi-edmft}
\end{eqnarray}
(More specifically, the spin self-energy is defined in terms of 
an effective spin cumulant that is
$I$-irreducible.\cite{Smith1})
The conduction-electron Green's function is, as usual,
expressed in terms of
the conduction-electron self-energy $\Sigma(\omega)$:
\begin{eqnarray}
G ({\bf k}, \omega) &=& \frac {1}  {\omega + \mu - \epsilon_{\bf k}
- \Sigma (\omega)} .
\label{g-edmft}
\end{eqnarray}
A
crucial
advantage of the EDMFT is that, unlike
standard RPA approximations,
the damping of spin fluctuations is not restricted
to take place via decay into
quasiparticle-quasihole pairs. Instead the (dynamical) solution
of the problem will dictate 
the nature of the low-energy many-body excitations and the associated 
form of the spin damping. At the same time,
it is important to bear in mind the
central
approximation made in
the EDMFT, namely, both $\Sigma$ and $M$ are taken to be momentum-independent.

\begin{figure}
\vspace{2ex}
\centering
\vbox{\epsfxsize=65mm\epsfbox{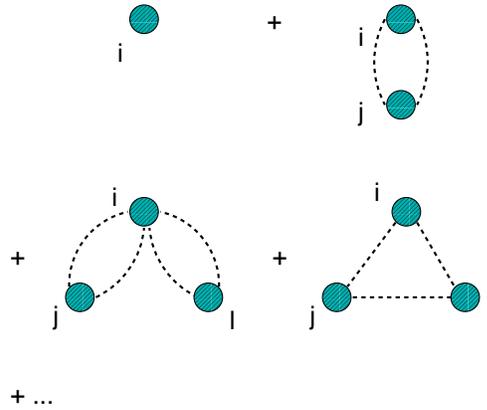}}
\vspace{3ex}
\caption{Single-site, two-site, and three-site diagrams for the
Luttinger-Ward potential in the extended dynamical mean-field theory.
A dashed line represents an intersite interaction;
this is $I_{ij}$ for the Hamiltonian given in Eq.~(\ref{kondo-lattice}).
A blob contains all the on-site diagrams, with the associated
(fully dressed) fermion Green's function.
The series extends to diagrams involving an infinite number of sites.}
\label{edmft-diagrams}
\end{figure}

Details of the formalism have been published
in Ref.~\onlinecite{Smith1}. Here we stress that the EDMFT can be
adopted as a conserving resummation of diagrams for finite-dimensional
systems. As discussed in Ref.~\onlinecite{Smith1}, the diagrams
retained in the EDMFT form an infinite series, which is
illustrated in Fig.~\ref{edmft-diagrams}.
(Note that standard Feynman diagrams can be used for the Kondo lattice
model once the spin of the local moment is represented in terms of
pseudo-fermions without a constraint; one such representation suitable
for the spin-${1 \over 2}$ case we are considering is the
Popov representation.\cite{Popov})

The spatial dimensionality $d$
and other aspects of the lattice structure are encoded
in the form of the tight-binding parameter $t_{ij}$ and the RKKY
interaction $I_{ij}$. In particular, a quantity that will play a crucial
role in our analysis is the ``RKKY density of states,''
\begin{eqnarray}
\rho_{I} (\epsilon) \equiv  \sum_{\bf q} \delta
( \epsilon  - I_{\bf q} ) .
\label{rkky-dos}
\end{eqnarray}

For any finite-dimensional system, $I_{\bf q}$ is nonzero only over
a finite range. In other words, the support of $\rho_{I} (\epsilon)$
is bounded. This ensures that there is a finite region of parameter
space over which $\chi ({\bf q}, \omega)$ is positive and nonsingular;
hence, a stable paramagnetic solution exists. An alternative construction of
the EDMFT, also given in Ref.~\onlinecite{Smith1}, scales the intersite
parameters such that the nearest-neighbor coupling goes as
$I_{\langle ij\rangle} \propto 1/\sqrt{d}$, and then takes the limit
$d \rightarrow \infty$. In this case, a special choice of lattice has
to be made in order to obtain a stable paramagnetic solution.

\subsection{Extended dynamical mean-field equations for the Kondo lattice}

The EDMFT equations for the Kondo lattice model can be expressed in terms of
an effective action for a single lattice site (the ``impurity'' site):
\begin{eqnarray}
{\cal S}_{\text{imp}} = && {\cal S}_{\text{top}}
+\int_0^{\beta} d \tau ~J_K ~{\bf S} \cdot {\bf s}_{c} \nonumber\\
&&- \int_{0}^{\beta} d \tau
\int_{0}^{\beta} d \tau' \left[ \sum_{\sigma} c_{\sigma} ^ {\dagger} (\tau)
G_0^{-1}(\tau - \tau ') c_{\sigma}(\tau') \right. \nonumber\\
&& \left. + \; 
{1 \over 2}
{\bf S}(\tau) \cdot \chi_{0}^{-1}(\tau - \tau') {\bf S}(\tau')
\rule[-2ex]{0ex}{4ex}\right] ,
\label{S-imp-kondo-lattice}
\end{eqnarray}
where $\beta = 1/k_B T$, ${\cal S}_{\text{top}}$ describes the
Berry phase
of the local moment, and $G_0^{-1 } $ and $\chi_{0}^{-1}$ are Weiss fields.

Equivalently, the effective impurity action can be represented in terms
of the following effective impurity Hamiltonian:
\begin{eqnarray}
{\cal H}_{\text{imp}}
=&& J_K ~{\bf S} \cdot {\bf s}_c
+ \sum_{p,\sigma} E_{p}~c_{p\sigma}^{\dagger}~ c_{p\sigma}
\nonumber\\ &&
+ \; g \sum_{p} {\bf S} \cdot
\left( \vec{\phi}_{p} + \vec{\phi}_{-p}^{\;\dagger} \right)
+ \sum_{p} w_{p}\,\vec{\phi}_{p}^{\;\dagger}\cdot \vec{\phi}_{p} .
\label{H-imp}
\end{eqnarray}
This is the Bose-Fermi Kondo Hamiltonian, describing a local moment coupled
not only to the spin of a fermionic bath ($c_{p\sigma}$), as in the
usual Kondo problem, but also to a dissipative vector-bosonic bath
($\vec{\phi}_{p}$). Physically, the vector bosons describe a
fluctuating magnetic field generated via the RKKY interaction
by the local moments at all
other lattice sites.
The dispersions, $E_p$ and $w_p$, as well as the coupling constant $g$,
are such that integrating out the two baths in the Hamiltonian
representation yields the
effective action~(\ref{S-imp-kondo-lattice}). Specifically,
\begin{eqnarray}
g^{2} \sum_{p}  { 2 w_{p} \over {(i\nu_m)^{2} - w_{p}^{2}}}
&=& - \chi_{0}^{-1}(i\nu_m),
\nonumber \\[-1ex]
\label{parameter_definitions} \\[-1ex]
\sum_{p} { 1 \over {i\omega_n - E_p}} &=& G_0(i \omega_n) ,
\nonumber
\end{eqnarray}
where $\nu_m$ and $\omega_n$ are
bosonic and fermionic Matsubara frequencies, respectively.
In the following, we will interchange between
the two representations based on notational convenience.

The self-consistent procedure goes as follows:
\begin{enumerate}
\item
Put in trial forms for
$\chi_0^{-1} (i \nu_m )$
and
$G_0^{-1}( i \omega_n)$.
\item
Solve the impurity Kondo problem to determine the impurity
correlation functions,
\begin{eqnarray}
\chi_{\text{imp}} (\tau) \equiv \langle T_{\tau} S^x (\tau) S^x (0)\rangle
\label{chi-loc}
\end{eqnarray}
and
\begin{eqnarray}
G_{\text{imp}} (\tau) \equiv -\langle T_{\tau} c_{\sigma} (\tau)
c_{\sigma}^{\dagger} (0)\rangle .
\label{G-loc}
\end{eqnarray}

This step also determines the spin self-energy
through a Dyson-like equation for the impurity problem,
\begin{eqnarray}
M_{\text{imp}}
( i \nu_m ) = \chi_{0}^{-1}( i \nu_m ) + {1 \over \chi_{\text{imp}}
(i \nu_m) } ,
\label{spin-se}
\end{eqnarray}
and a conduction-electron self-energy,
\begin{eqnarray}
\Sigma_{\text{imp}} (i \omega_n) = G
_{0}^{-1}(i \omega_n) - {1 \over G_{\text{imp}}(i \omega_n)} .
\label{electron-se}
\end{eqnarray}
\item
Determine the parameters in the trial forms of $\chi_0^{-1} (i \nu_m )$ and
$G_0^{-1}( i \omega_n) $ by demanding self-consistency.
The self-consistency conditions
\begin{eqnarray}
\chi_{\text{imp}} (\omega) &=& \sum_{\bf q} \chi ( {\bf q}, \omega ) ,
\nonumber \\[-1ex]
\label{self-consistent} \\[-1ex]
G_{\text{imp}} (\omega) &=& \sum_{\bf k} G( {\bf k}, \omega ) ,
\nonumber
\end{eqnarray}
amount to the requirement that each local correlation function is equal to
the wave-vector average of the corresponding lattice correlation function.
The lattice spin susceptibility and lattice Green's function,
$\chi ( {\bf q}, i \nu_m )$ and $G( {\bf k}, i \omega_n )$, respectively,
are given in Eqs.~(\ref{chi-edmft}) and~(\ref{g-edmft}).

\item
Once self-consistency is achieved,
identify the impurity quantities $\chi_{\text{imp}}$ and $G_{\text{imp}}$
with the corresponding local (on-site) quantities $\chi_{\text{loc}}$
and $G_{\text{loc}}$ of the lattice problem.
Likewise,
identify
$M_{\text{imp}}$ and $\Sigma_{\text{imp}}$
with the corresponding lattice quantities
$M$ and $\Sigma$,
and hence
use Eqs.~(\ref{chi-edmft}) and~(\ref{g-edmft}) to calculate
the lattice dynamical spin susceptibility $\chi ( {\bf q}, i \nu_m )$
and the lattice Green function $G( {\bf k}, i \omega_n )$.
\end{enumerate}

The derivation of Eqs.~(\ref{chi-edmft}), (\ref{g-edmft}), and~(\ref{spin-se})
has been given in Ref.~\onlinecite{Smith1}.
For notational simplicity, in the remainder of the paper we will
not differentiate between
impurity quantities and local
quantities. In particular, the impurity spin-spin correlation function
determined from the Bose-Fermi Kondo model will simply be called
$\chi_{\text{loc}}$.

\section{Solution of the impurity problem}
\label{sec:impurity}

Our goal is to determine the universal low-energy behavior
of the effective impurity problem generated by the EDMFT.
For this purpose, we choose trial forms for the Weiss fields
such that the density of states of the fermion
bath near the Fermi energy is a nonzero constant,
\begin{equation}
\sum_{p} \delta (\omega - E_{p}) = N_0 ,
\label{dos-fermion}
\end{equation}
while the spectral function of the fluctuating magnetic field
has a sublinear power-law dependence on energy at sufficiently low energies:
\begin{eqnarray}
\sum_{p}  \delta (\omega - w _p)  = 
(K_0^2 / \pi) |\omega|^{\gamma} ~~~~~~~~\text{for} ~ |\omega| < \Lambda ,
\label{dos-boson}
\end{eqnarray}
with $\gamma < 1$.
Equivalently---through Eq.~(\ref{parameter_definitions})---the
imaginary part of the Weiss field takes the form
\begin{equation}
\text{Im} \, \chi_0^{-1} (\omega + i0^+) = C |\omega|^{\gamma}
\text{sgn}\, \omega ,
\label{dos-boson2}
\end{equation}
where
\begin{equation}
C = (K_0 g)^2 .
\label{C-def}
\end{equation}
Eq.~(\ref{dos-boson}) also defines the
cutoff parameter $\Lambda$.
The self-consistency of our
trial forms for the Weiss fields
will be established in Section~\ref{sec:self-consistent},
where the solution for $\gamma$, $C$, $\Lambda$, and $N_0$ will also be given.

Having specified the parameters of the impurity model,
we now proceed to solve this model. Our strategy
is first to construct the RG equations
for the coupling constants $J_K$ and $g$. We then show that a phase
transition exists as the ratio $g/T_K^0$ is varied.
[Here we parametrize the Kondo coupling $J_K$ using
the single-impurity Kondo temperature,
\begin{eqnarray}
T_K^0 \approx {1 \over \rho_0 (\mu)} e^{-1/\rho_0 (\mu)J_K},
\label{TK}
\end{eqnarray}
where $\rho_0(\mu)$ is the bare conduction-electron density of states
at the chemical potential $\mu$.]
The critical value
of $g/T_K^0$ is determined to linear order in $1-\gamma$.
We then calculate the correlation functions at the critical point,
also to linear order in $1-\gamma$.
This $(1\!-\!\gamma)$-expansion
was first introduced to the Bose-Fermi Kondo model
by Smith and Si\cite{Smith3}
[following an earlier
$(1\!-\!\gamma)$-expansion for a spinless Bose-Fermi
Kondo model\cite{Smith2}] and by Sengupta.\cite{Sengupta}
It was subsequently 
used to study
a Bose Kondo model with an interacting bath 
by Vojta, Buragohain, and Sachdev.\cite{Vojta}

Both the construction of the RG equations and the calculation of
correlation functions require a proper handling of the spin operators.
Following Smith and Si,\cite{Smith3}
we adopt the Abrikosov representation of the
spin in terms of pseudo-$f$-electrons,\cite{Abrikosov}
\begin{equation}
{\bf S} = \sum_{\sigma \sigma'} f_{\sigma}^{\dagger}
{\vec{\tau}_{\sigma\sigma'} \over 2} f_{\sigma'} ,
\label{pseudo-fermion}
\end{equation}
where $\tau^{x,y,z}$ are the Pauli matrices. To stay within the Hilbert
space for the spin, we assign an energy $\lambda$ to the $f$-electron
level and calculate any correlation function (${\cal D}$)
involving the spin from the corresponding
correlation function in the $f$-electron representation
($\tilde{\cal D}$) using
\begin{eqnarray}
{\cal D} = \lim_{\lambda \rightarrow \infty}
{1 \over 2} ~ e^{\beta \lambda}~ \tilde{\cal D} .
\label{cf-abrikosov}
\end{eqnarray}
Here the prefactor ${1 \over 2}\,e^{\beta \lambda}$ is introduced
to compensate the Boltzmann weighting factor for the singly occupied
subspace in the pseudo-fermion basis.
Wick's theorem applies to $\tilde{\cal D}$
since it involves only canonical fermion operators.

\begin{figure}
\vspace{2ex}
\centering
\vbox{\epsfxsize=60mm\epsfbox{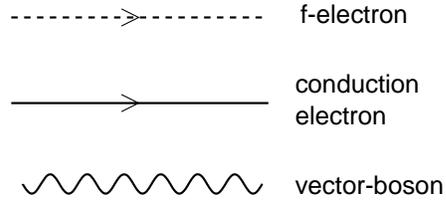}}
\vspace{3ex}
\caption{Bare propagators of the impurity problem
defined by the Hamiltonian~(\ref{H-imp}), with the spin
of the local moment represented in terms of
(pseudo-)$f$-electrons.}
\label{bare-propagators}
\end{figure}

Fig.~\ref{bare-propagators}
shows the diagrammatic representation of
the bare propagators:
\begin{equation}
G_f^b ( i\omega_n ) = { 1 \over {i\omega_n - \lambda} }
\label{Gf-bare}
\end{equation}
for the $f$-electron,
\begin{equation}
G^b ({p}, i\omega_n ) = { 1 \over {i\omega_n - E_{p}}}
\label{G-bare}
\end{equation}
for the conduction electrons,
and
\begin{equation}
G_{\phi}^b (p, i \nu_m ) = { 2 w_p \over {(i \nu_m)^2 - w_p^2 }}
\label{G-phi-bare}
\end{equation}
for the vector bosons.
$G_{\phi}^b (p, i \nu_m )$ is the Fourier transform
of $G_{\phi}^b (p, \tau )
\equiv -\langle T_{\tau} (\vec{\phi}_p\!+\!\vec{\phi}_{-p}^{\;\dagger}) (\tau)
\!\cdot\!(\vec{\phi}_p\!+\!\vec{\phi}_{-p}^{\;\dagger})(0) \rangle/3$.

\begin{figure}
\vspace{2ex}
\centering
\vbox{\epsfxsize=60mm\epsfbox{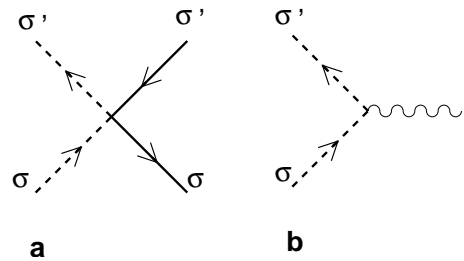}}
\vspace{3ex}
\caption{(a) Kondo coupling between the spin of
an
$f$-electron
and the spin of a conduction electron; (b)~coupling between
the spin of
an
$f$-electron and the vector-boson field.}
\label{couplings}
\end{figure}

In the pseudo-fermion representation, the Hamiltonian terms
describing the Kondo interaction and the local moment-vector boson
interaction take the forms
\begin{eqnarray}
H_{J} &=& {J_K \over 4} \sum_{\sigma \sigma'} \sigma \sigma'
f_{\sigma}^{\dagger} f_{\sigma} c_{\sigma'}^{\dagger}
c_{\sigma'}
+{J_K \over 2} ( f_{\uparrow}^{\dagger} f_{\downarrow}
c_{\downarrow}^{\dagger}
c_{\uparrow} + \text{H.c.}),
\nonumber \\[-1ex]
\label{coupling} \\[-1ex]
H_{g} &=& {g \over 2} \sum_{\sigma } \sigma
f_{\sigma}^{\dagger} f_{\sigma} {\phi}^z
+{g \over \sqrt{2}} ( f_{\uparrow}^{\dagger} f_{\downarrow}
{\phi}^{-} + \text{H.c.}) ,
\nonumber
\end{eqnarray}
where
$\sigma, \sigma'
= \pm 1$,
$\vec{\phi} = \sum_p (\vec{\phi}_p + \vec{\phi}_{-p}^{\;\dagger})$,
and ${\phi}^{\pm} = (\phi^x \pm \phi^y)/\sqrt{2}$.
$H_J$ and $H_g$ are shown graphically in Fig.~\ref{couplings}.

\subsection{RG equations}
\label{sec:rg}

We first determine the bare scaling dimensions of the couplings $J_K$ and $g$.
${\bf S}(\tau)$ is dimensionless because when $J_K=g=0$,
$\langle{\bf S}(\tau) \cdot {\bf S}(0)\rangle_0$ is a constant.
$\vec{\phi}(\tau)$ has the scaling dimension of 
$\tau^{-(1 + \gamma)/2}$, and 
${\bf s}_c(\tau)$ that of $\tau^{-1}$.
Therefore, $J_K$ is marginal, while $g$ has scaling dimension
$(1-\gamma)/2$, i.e., $g$ is marginal for $\gamma=1$ and relevant
for $\gamma < 1$.

We now calculate the $\beta$ function to cubic order in
the coupling constants. To do so, we first calculate the perturbative
contributions to the $f$-electron self-energy, and the perturbative
corrections to the vertices shown in Fig.~\ref{couplings}.
(The self-energies of the fermionic and bosonic baths are
of order $1/N_{\rm site}$ and vanish in the thermodynamic
limit for the baths or, equivalently, when the
bath spectra
are continuous.)
Details of the calculation
are given in Appendix~\ref{sec:appen-rg}.

\begin{figure}
\vspace{2ex}
\centering
\vbox{\epsfxsize=40mm\epsfbox{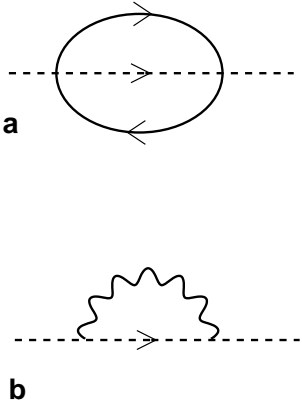}}
\vspace{3ex}
\caption{Diagrams for the $f$-electron self-energy.}
\label{f-self-energy}
\end{figure}

Consider first the $f$-electron self-energy. The relevant diagrams
for the perturbative corrections are
given in Fig.~\ref{f-self-energy}. 
The result is:
\begin{eqnarray}
\Sigma_f(\omega + \lambda )
= - \left[{3 \over 8} (N_0J_K )^2 + {3 \over 4} (K_0g)^2 \right]
\omega \ln {W \over \omega} ,
\label{Sigma-f}
\end{eqnarray}
where $W$ is a running cutoff energy. Introducing ${\cal G}_f$ through
\begin{eqnarray}
G_f(\omega + \lambda ) = G_f^{\text{b}} (\omega + \lambda )
{\cal G} _f (\omega) ,
\label{cal-G-f-def}
\end{eqnarray}
we have
\begin{eqnarray}
{\cal G} _f (\omega)
= 1 - \left[{3 \over 8} (N_0J_K )^2 + {3 \over 4} (K_0g)^2 \right]
\ln {W \over \omega} .
\label{cal-G-f}
\end{eqnarray}

\begin{figure}[t]
\vspace{2ex}
\centering
\vbox{\epsfxsize=75mm\epsfbox{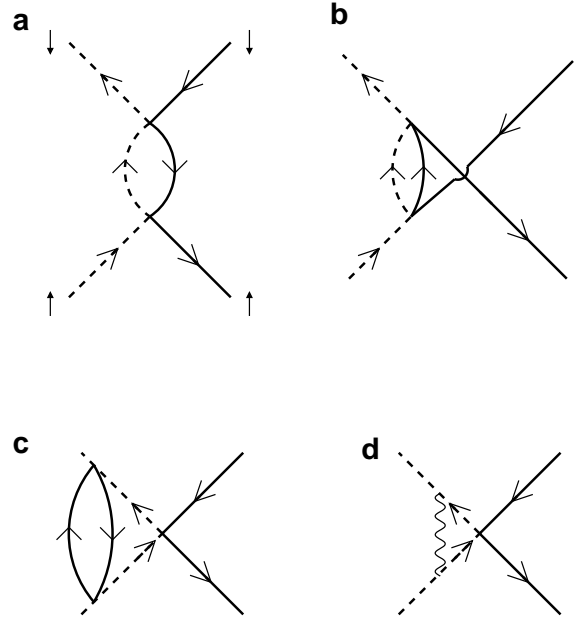}}
\vspace{3ex}
\caption{Perturbative corrections to the Kondo coupling vertex.}
\label{J-vertex}
\end{figure}

Next consider the corrections to the Kondo coupling $J_K$.
Due to spin-rotational invariance it suffices to
examine just the spin-flip processes, say.
Also, we set the energy
of the external $f$-electrons to $\omega+\lambda$ and that of
the external conduction electrons to $\mu$.
The diagrams that lead to singular corrections to the full
vertex $\Gamma_J(\omega)$ are given in Fig.~\ref{J-vertex}.
We write the full vertex as
\begin{eqnarray}
\Gamma_J (\omega ) = J_K \gamma_J (\omega ),
\label{gamma-J-def}
\end{eqnarray}
and find that
\begin{eqnarray}
\gamma_J(\omega)
= 1 + \left[ N_0J_K \! - \! {1 \over 8} (N_0 J_K)^2 \! - \! { 1 \over 4}
(K_0 g)^2 \right] \ln {W \over \omega} .
\label{gamma-J}
\end{eqnarray}

\begin{figure}[t]
\vspace{2ex}
\centering
\vbox{\epsfxsize=65mm\epsfbox{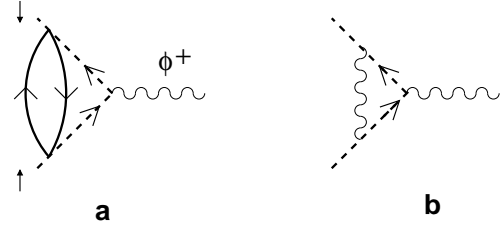}}
\vspace{3ex}
\caption{Perturbative corrections to the vertex describing the coupling
of the local moment to the vector bosons.}
\label{g-vertex}
\end{figure}

Finally, consider the corrections to the coupling constant $g$.
The diagrams contributing
singular corrections to the
corresponding full vertex $\Gamma_g $
are given in Fig.~\ref{g-vertex}. Writing
\begin{eqnarray}
\Gamma_g (\omega ) = g ~\gamma_g (\omega ),
\label{gamma-g-def}
\end{eqnarray}
we have
\begin{eqnarray}
\gamma_g(\omega)
= 1 - \left[ {1 \over 8} (N_0 J_K)^2 + { 1 \over 4}
(K_0 g)^2 \right] \ln {W \over \omega} .
\label{gamma-g}
\end{eqnarray}

We are now in a position to construct the RG equations.
Lowering the cutoff energy from $W$ to $W'$ produces
the
coupling-constant and wave-function renormalizations
\begin{eqnarray}
{\cal G}_f (\omega, W', J_K', g') &=&
z_f \, {\cal G}_f (\omega, W, J_K, g) ,
\nonumber\\[1ex]
\gamma_J (\omega, W', J_K', g') &=&
z_J^{-1} \gamma_J (\omega, W, J_K, g) ,
\label{z} \\[1ex]
\gamma_g (\omega, W', J_K', g') &=&
z_g^{-1} \gamma_g (\omega, W, J_K, g) ,
\nonumber
\end{eqnarray}
where
\begin{equation}
J_K' = z_f^{-1}z_J J_K, \quad
g' = z_f^{-1}z_g g .
\label{j'}
\end{equation}
This leads to the RG equations
\begin{eqnarray}
{{ d J_K} \over { d l}}
&& = J_K \left[ N_0J_K - {1 \over 2} N_0^2 J_K^2
- K_0^2  g^2 \right] ,
\nonumber \\[-1ex]
\label{rg-eq} \\[-1ex]
{{ d g} \over  {d l}}
 && = g \left[ {{1 - \gamma} \over 2} -
{1 \over 2} N_0^2 J_K^2 - K_0^2  g^2 \right] ,
\nonumber
\end{eqnarray}
where $l =  \ln(W/W')$.

\begin{figure}
\vspace{2ex}
\centering
\vbox{\epsfxsize=50mm\epsfbox{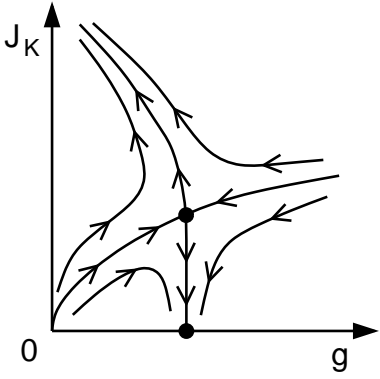}}
\vspace{3ex}
\caption{Schematic RG flow of the Bose-Fermi Kondo model.
Note the unstable fixed point at finite $J_K$ and $g$.}
\label{rg-flow}
\end{figure}

To linear order in $1-\gamma$,
the $\beta$ functions given in Eq.~(\ref{rg-eq})
yield an unstable
fixed point located in the $J_K$-$g$ plane at
\begin{equation}
(K_0 g^*)^2 =
N_0 J_K^* = (1-\gamma)/2 .
\label{fp}
\end{equation}
A separatrix passes through this critical point and through the origin,
$J_K = g = 0$: to its left (right) in the $J_K$-$g$ plane
the RG flow of $J_K$ is towards infinity (zero)
(see Fig.~\ref{rg-flow}).
For small $J_K$ and $g$, the
position of the separatrix, expressed in the form $g_c(J_K)$,
can be determined by iterating the RG flow.
This calculation, the details of which appear in
Appendix~\ref{sec:appen-separatrix}, gives
\begin{eqnarray}
(K_0 g_c )^2 = 2 (1 - \gamma) \exp\left(-{{1-\gamma} \over N_0J_K }\right).
\label{separatrix}
\end{eqnarray}

\subsection{Correlation functions at the critical point}
\label{sec:cf}

Consider first the local spin autocorrelation function
\begin{eqnarray}
\chi_{\text{loc}}(\tau) \equiv \langle T_{\tau} S^x (\tau) S^x (0) \rangle
= \lim_{\lambda \rightarrow \infty} {1 \over 2}~ e^{\beta \lambda}~
\tilde {\chi}(\tau) ,
\label{chi-tau}
\end{eqnarray}
where
\begin{eqnarray}
\tilde {\chi}
(\tau) \equiv {1 \over 2} \langle T_{\tau}
f_{\downarrow}^{\dagger}(\tau) f_{\uparrow} (\tau)
f_{\uparrow}^{\dagger}(0) f_{\downarrow} (0)\rangle ,
\label{chi-tilde}
\end{eqnarray}
which can be calculated using diagrammatic techniques.
The correlation functions along the entire separatrix
specified by Eq.~(\ref{separatrix}) are asymptotically the same
as their counterparts at the fixed point described by Eq.~(\ref{fp}).

\begin{figure}[t]
\vspace{2ex}
\centering
\vbox{\epsfxsize=65mm\epsfbox{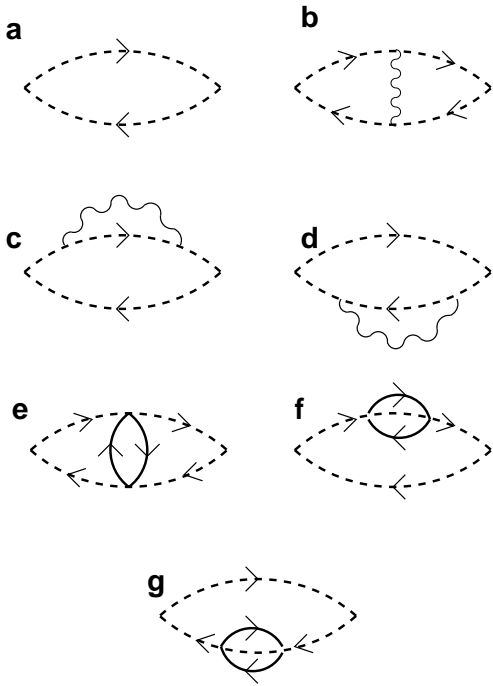}}
\vspace{3ex}
\caption{Perturbation series for $\tilde{\chi}_{\text{loc}}$, defined
in Eq.~(\ref{chi-tilde}), showing terms entering at zeroth (a),
first (b)--(d), and higher (e)--(g) order in $1-\gamma$.}
\label{chi-diagrams}
\end{figure}

As usual, given that $1-\gamma$ is
treated as an expansion parameter, the correlation functions
at the critical point can be calculated
via
perturbative expansion
in the couplings $J_K$ and $g$, at the end
replacing these couplings by their fixed-point values,
$J_K^*$ and $g^*$, respectively. The low-order diagrams are
shown in Figs.~\ref{chi-diagrams}(a)--(g).

Because $J_K^*$ is of order $1-\gamma$, the corrections to
${\chi}_{\text{loc}}$ from the Kondo coupling $J_K$
[Figs.~\ref{chi-diagrams}(e), \ref{chi-diagrams}(f),
and~\ref{chi-diagrams}(g)]
are of higher-than-linear order in $1-\gamma$.
The calculation of the remaining contributions is
given in detail in Appendix~\ref{sec:appen-corr}.
Diagram~\ref{chi-diagrams}(a) contributes to zeroth order in
$1-\gamma$:
\begin{eqnarray}
\chi_{\text{loc}}^{(a)}(\tau) = 1 / 4 .
\label{chi-0}
\end{eqnarray}
Three diagrams contribute to first order in $1-\gamma$.
The singular correction from diagram~\ref{chi-diagrams}(b) is
\begin{eqnarray}
\chi_{\text{loc}}^{(b)}(\tau) = -{1 \over 8} (K_0 g^*)^2
\ln { { \sin ( \pi \tau / \beta) } \over
{\pi / \Lambda \beta } } .
\label{chi-b}
\end{eqnarray}
(Here we consider $0<\tau < \beta$.)
The singular contributions from diagrams~\ref{chi-diagrams}(c) and
\ref{chi-diagrams}(d) sum to
\begin{eqnarray}
\chi_{\text{loc}}^{(cd)}(\tau) = -{3 \over 8} (K_0 g^*)^2
\ln { { \sin ( \pi \tau / \beta )} \over
{\pi /\Lambda \beta }  } .
\label{chi-cd}
\end{eqnarray}
Collecting all these contributions, noting that
$(K_0g^*)^2 = (1 -\gamma)/2$, and
following the procedure of the standard
$\epsilon$-expansion,\cite{Wilson-Kogut}
we end up with the following form for the
local spin correlation function at the critical point:
\begin{eqnarray}
\chi_{\text{loc}}(\tau) = {1 \over 4} \left (
{ {\pi / \Lambda \beta }
\over
{ \sin ( \pi \tau / \beta)}
} \right )^{1 -\gamma} .
\label{chi}
\end{eqnarray}

In the zero-temperature limit, this gives,
for $0 < \gamma < 1$,
\begin{equation}
\chi_{\text{loc}} (\omega ) = { 1 \over 2 \Lambda^{1-\gamma}}
\Gamma(\gamma)
\sin{{\pi (1 - \gamma)}
\over 2} (-i\omega)^{-\gamma} ,
\label{chi-loc-epsilon-omega}
\end{equation}
where $\Gamma$ is the gamma function, and for
$\gamma = 0$,
\begin{equation}
\chi_{\text{loc}} (\omega ) = { 1 \over 2 \Lambda }
\ln { {\Lambda} \over {- i \omega}} .
\label{chi-loc-epsilon-omega-gamma=0}
\end{equation}

\begin{figure}[t]
\vspace{2ex}
\centering
\vbox{\epsfxsize=45mm\epsfbox{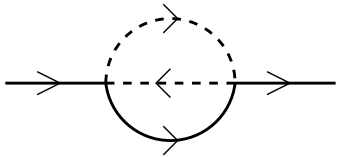}}
\vspace{3ex}
\caption{Leading nonvanishing contribution to the
conduction-electron self-energy.}
\label{Sigma-c-diagrams}
\end{figure}

Now we turn to
the conduction-electron Green's function.
The leading nonvanishing contribution to the conduction-electron
self-energy is given in Fig.~\ref{Sigma-c-diagrams}.
Since $J_K^*$ is of order $1-\gamma$, we conclude that
\begin{equation}
\Sigma = O((1-\gamma)^2) ,
\label{Sigma-c-qcp}
\end{equation}
i.e., the conduction-electron Green's function is unrenormalized
to linear order in $1-\gamma$.
Note that this stands in marked contrast to the strong-coupling
Kondo phase, where $\Sigma$ has a pole.\cite{Hewson}

\begin{figure}
\vspace{2ex}
\centering
\vbox{\epsfxsize=65mm\epsfbox{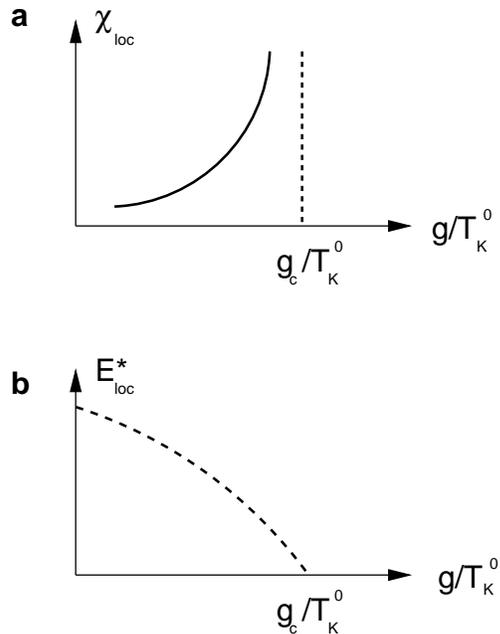}}
\vspace{3ex}
\caption{(a)
Static local
spin susceptibility as a function of
$g/T_K^0$. For $g < g_c$, the susceptibility is finite and the
local moment is completely quenched. The divergence of the susceptibility
at $g=g_c$ signals that the dynamics of the local moment have become
critical.
(b) The local energy scale $E_{\text{loc}}^*$ as a function of
$g/T_K^0$. $E_{\text{loc}}^*$ vanishes at $g_c$, reflecting a
critical slowing down.}
\label{impurity-qcp}
\end{figure}

To summarize, for a fixed $J_K$, the impurity model displays a
QCP at a critical coupling, $g_c(J_K)$
specified by Eq.~(\ref{separatrix}).
The static impurity spin susceptibility
diverges as $g\rightarrow g_c^-$,
as illustrated in Fig.~\ref{impurity-qcp}(a).
At the
QCP,
the dynamical susceptibility is given by Eq.~(\ref{chi-loc-epsilon-omega}).
Associated with this critical point is the vanishing
of a local energy scale $E_{\text{loc}}^*$, reflecting a critical
slowing down. This is
shown schematically in Fig.~\ref{impurity-qcp}(b).
For $g < g_c$, $E_{\text{loc}}^*$ is
finite and serves as an effective Fermi energy scale. 

Subsequent to the work reported here, the RG analysis of the Bose-Fermi
Kondo model has been extended\cite{Zhu:02,Zarand:02} to higher orders
in $1-\gamma$.
An unstable fixed point has been identified not only in the spin-isotropic
model considered above, but also in the presence of anisotropy.
In both cases, the exponent of the local spin correlation function at the
critical point turns out to be $1-\gamma$ to all orders in $1-\gamma$,
in agreement with Eq.~(\ref{chi}).

We note that power-law behavior of local spin-correlation functions 
has also been discussed in impurity models that arise in the context of
quantum spin-glasses.\cite{Sachdev-Ye93,Sengupta,Grempel,Parcollet}
There, it is associated with a {\it stable} fixed point (i.e., a phase)
of the impurity model.
By contrast, what
we have discussed above is an
{\it unstable} fixed point (a critical point).

\section{Self-consistent solution}
\label{sec:self-consistent}

We must now address two key questions:
\begin{itemize}
\item Can the critical point associated with the local problem
coincide with the magnetic ordering transition?
\item What can cause the fluctuating magnetic field to have a
sub-ohmic spectrum ($\gamma < 1$)?
\end{itemize}

The answer to both questions lies in the self-consistency
requirement.
The determining factor turns out to be the form of the
RKKY density of states defined in Eq.~(\ref{rkky-dos}).
The significance of $\rho_I(\epsilon)$ is that it
characterizes the way the RKKY interaction enters
the self-consistency condition. Indeed,
combining Eq.~(\ref{rkky-dos}) with
Eq.~(\ref{chi-edmft}), we can rewrite the first equation in
the self-consistency condition~(\ref{self-consistent}) as
\begin{eqnarray}
\chi_{\text{loc}} (\omega) &=&
\int d \epsilon
\frac {\rho_I (\epsilon )}
{ M(\omega) + \epsilon } .
\label{self-consistent-2}
\end{eqnarray}

\begin{figure}
\vspace{2ex}
\centering
\vbox{\epsfxsize=55mm\epsfbox{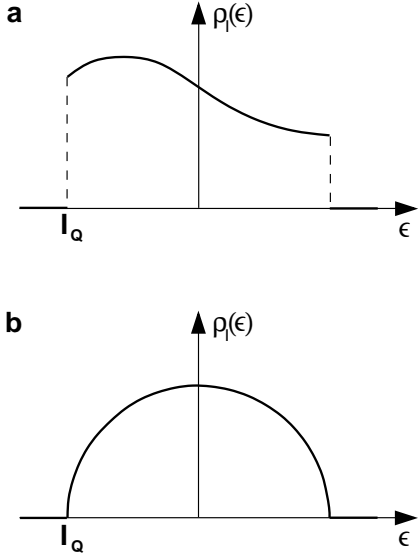}}
\vspace{3ex}
\caption{(a) An RKKY density of states with a jump at the lower
edge, $\epsilon = I_{\bf Q}$, characteristic of two-dimensional
magnetic fluctuations; (b) An RKKY density of states with a square-root
onset at the lower edge, characteristic of
three-dimensional magnetic fluctuations.}
\label{rkky-dos-2d3d}
\end{figure}

We consider two types of RKKY density of states $\rho_{I}(\epsilon)$:
The first type increases from the lower
band edge (at $\epsilon = I_{\bf Q}$)
with a jump, as
shown
in Fig.~\ref{rkky-dos-2d3d}(a).
The second type increases from the lower band edge
in a square-root fashion, i.e., as $\sqrt{\epsilon - I_{\bf Q}}\,$.
This is illustrated in Fig.~\ref{rkky-dos-2d3d}(b).

The jump and square-root onset at the lower edge are characteristic of
magnetic fluctuations in two dimensions and three dimensions,
respectively. This can be seen,
for instance, by comparing the form of the RKKY density of states
for nearest-neighbor real-space coupling on a
(two-dimensional) square lattice, where
$I_{\bf q} = I (\cos q_x a + \cos q_y a)$,
and a (three-dimensional) cubic lattice, where
$I_{\bf q} = I (\cos q_x a + \cos q_y a + \cos q_z a )$.

In the remainder of this section,
we describe the quantum critical behavior
for the two types of RKKY density of states.
The quantum phase transition takes place as
the ratio of the RKKY interaction to the Kondo coupling,
\begin{eqnarray}
\delta \equiv {{|I_{\bf Q}|} \over T_K^0},
\label{delta}
\end{eqnarray}
is tuned through some
critical
value $\delta_c$, which labels the QCP.
We expect $\delta_c$, the ratio of energy scales characterizing the competing
interactions that give rise to the transition, to be of order unity.

The QCP is signaled by the divergence of the static
spin susceptibility at the peak wavevector ${\bf Q}$.
The peak susceptibility is related to the spin self-energy $M(\omega)$
via Eq.~(\ref{chi-edmft}):
\begin{eqnarray}
\chi({\bf Q}, \omega) = \frac {1}  { M(\omega) + I_{\bf Q} } .
\label{chi-Q}
\end{eqnarray}
This implies that at the QCP,
\begin{eqnarray}
M(\omega\rightarrow 0) = - I_{\bf Q} .
\label{M-QCP}
\end{eqnarray}

\subsection{Self-consistent solution in two dimensions}
\label{sec:self-consistent2D}

With a jump in the RKKY density of states, the self-consistency
equation Eq.~(\ref{self-consistent-2})
can be written as
\begin{eqnarray}
\chi_{\text{loc}} (\omega) = \int_{I_{\bf Q}}^{I_{\bf Q}+\Lambda_0}
\!\!\! d \epsilon \frac {\rho_I (\epsilon) }
  { M(\omega) + \epsilon }
+ \int_{I_{\bf Q}+\Lambda_0} \!\!\! d \epsilon
\frac {\rho_I (\epsilon) } { M(\omega) + \epsilon } ,
\label{self-consistent-2D-sep}
\end{eqnarray}
where $\Lambda_0$ is defined
such that,
for $\epsilon \in (I_{\bf Q}, I_{\bf Q}+\Lambda_0)$,
$\rho_I (\epsilon)$ is approximately equal to its value
at the lower edge, $\rho_I (I_{\bf Q})$.
For instance, if $\rho_I(\epsilon)$ is flat, then
$\Lambda_0$ is equal to the separation between the upper and lower
edge of the flat $\rho_I(\epsilon)$.
For a generic $\rho_I(\epsilon)$,
$\Lambda_0$ is of the order of the RKKY
interaction as parametrized by, say,
$|I_{\bf Q}|$.
Given Eq.~(\ref{M-QCP}),
the first term in Eq.~(\ref{self-consistent-2D-sep}) is
singular;
the second term, on the other hand, is nonsingular
and can be neglected. The self-consistency equation
then becomes
\begin{eqnarray}
\chi_{\text{loc}} (\omega) = \rho_I (I_{\bf Q}) \ln
{\Lambda_0 \over {M(\omega) + I_{\bf Q} }} .
\label{self-consistent2D}
\end{eqnarray}

Eq.~(\ref{self-consistent2D}),
when combined with Eqs.~(\ref{chi-Q}) and~(\ref{M-QCP}),
leads to the conclusion that the local susceptibility
$\chi_{\text{loc}} (\omega
\rightarrow 0
)$ becomes singular
at the same value of $\delta$ where the peak susceptibility
$\chi ({\bf Q}, \omega
\rightarrow 0
)$ becomes singular.
As usual, the singularity in $\chi ({\bf Q}, \omega)$
signals the emergence of a critical mode associated with the
long-wavelength fluctuations of the order parameter.
However, from Section~\ref{sec:cf}
[see the discussion concerning Fig.~\ref{impurity-qcp}(a)],
we know that the divergence of $\chi_{\text{loc}} (\omega)$ signifies
the emergence of critical but spatially local modes;
thus, the fluctuations of the local moments are also critical.
The corresponding energy scale
$E_{\text{loc}}^*$ vanishes [Fig.~\ref{impurity-qcp}(b)].

This reasoning leads to
one of our key conclusions concerning the quantum critical
behavior of the Kondo lattice in two
dimensions:
as we increase the parameter $\delta$, spatially local critical modes
emerge simultaneously with the usual long-wavelength critical modes.

We now determine the detailed dynamics at the
QCP, beginning with
the conduction-electron self-energy. Since the effective
impurity model is at its own critical point, our result of the
previous section [Eq.~(\ref{Sigma-c-qcp})] implies that
$\Sigma$
vanishes to linear order
in $1-\gamma$. This shows that the fermion-bath density of states,
assumed to be $N_0$ in Eq.~(\ref{dos-fermion}), is indeed a nonzero
constant and is, in fact, equal to the bare conduction-electron
density of states [see Eq.~(\ref{bare-cond-electron-dos})]
at the chemical potential:
\begin{eqnarray}
N_0 = \rho_0 (\mu).
\label{self-consistent-N0}
\end{eqnarray}

Let us
turn to the dynamical spin susceptibility.
Combining Eq.~(\ref{self-consistent2D}) with Eq.~(\ref{spin-se})
yields
\begin{eqnarray}
\chi_0^{-1} (\omega) + { 1 \over {\chi_{\text{loc}}(\omega)}}
&& = M(\omega) \nonumber \\
&&= - I_{\bf Q} + \Lambda_0 \exp
\left[ -{ {\chi_{\text{loc}}(\omega)} \over { \rho_I (I_{\bf Q})}} \right] .
\label{self-consistent2D-2}
\end{eqnarray}
Inserting Eqs.~(\ref{dos-boson2}), (\ref{separatrix}),
and~(\ref{chi-loc-epsilon-omega}) into Eq.~(\ref{self-consistent2D-2}),
and demanding a power-law form for the frequency-dependent part
of the spin self-energy $M(\omega)$, we reach the following self-consistent
solution for the parameters introduced in
Eqs.~(\ref{dos-boson})--(\ref{C-def}):
\begin{eqnarray}
\gamma = 0^+, \quad \Lambda = C /\pi = \frac{2}{\pi} T_K^0,
\label{solution-2D-parameters}
\end{eqnarray}
where $T_K^0$ is defined in Eq.~(\ref{TK}).
The specific form for the spectral function of the
vector-bosonic Weiss field is
\begin{eqnarray}
\text{Im} \chi_0^{-1} (\omega + i 0^+ ) =
\pi \Lambda \left ( \ln {\Lambda \over |\omega | }\right )^{-2}
\text{sgn}\, \omega ,
\label{solution-2D-weiss}
\end{eqnarray}
while the local susceptibility is
\begin{eqnarray}
\chi_{\text{loc}} (\omega + i 0^+ ) =  { 1 \over {2 \Lambda}}
\ln {\Lambda \over  {- i \omega }} .
\label{solution-2D-chi-loc}
\end{eqnarray}
The spin self-energy is given in terms of
$\chi_{\text{loc}}$ through the second equality of
Eq.~(\ref{self-consistent2D-2}).
The result is
\begin{eqnarray}
M (\omega ) =  - I _{\bf Q}
+ \Lambda_0 (-i\omega / \Lambda)^{\alpha} ,
\label{solution-2D-M}
\end{eqnarray}
where the exponent is
\begin{eqnarray}
\alpha =  { 1 \over { 2 \Lambda \rho_I (I_{\bf Q} )}} .
\label{solution-2D-alpha}
\end{eqnarray}
The parameter $\Lambda$ is determined by the Kondo coupling,
as shown in Eq.~(\ref{solution-2D-parameters}).
The parameter $\rho_I (I_{\bf Q} ) $, introduced earlier,
is the RKKY density of states at the lower edge
[Fig.~\ref{rkky-dos-2d3d}(a)].

\begin{figure}
\vspace{2ex}
\centering
\vbox{\epsfxsize=65mm\epsfbox{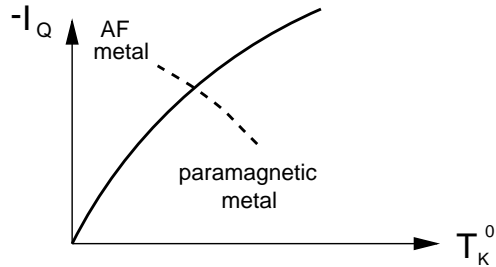}}
\vspace{3ex}
\caption{Phase diagram of the Kondo lattice in the
$|I_{\bf Q}|$-$T_K^0$ parameter space.
The solid line is the phase boundary. The dashed line shows
the tuning trajectory through the quantum phase transition for
a given system.}
\label{phase-diagram}
\end{figure}

We note that the local susceptibility at the critical point,
given in Eq.~(\ref{solution-2D-chi-loc}), is universal.
The exponent $\alpha$ for the spin self-energy, on the other hand,
depends on the product $\Lambda \rho_I (I_{\bf Q} )$.
Since the QCP is reached through competition between
the RKKY and Kondo interactions, we expect that this product
is close to unity, resulting in an $\alpha$ that is not too far
from ${1 \over 2}$. The precise value of this product,
however, is nonuniversal: it depends on which point of the
phase boundary in the two-dimensional
($|I_{\bf Q}|$-$T_K^0$) parameter space is crossed
as the system is tuned through the quantum phase transition
(see Fig.~\ref{phase-diagram}).

The phase boundary can, in principle, be located by inserting Eq.~(\ref{M-QCP})
into Eq.~(\ref{self-consistent2D-2}).
Recognizing that $1/\chi_{\text{loc}}(\omega \rightarrow 0) = 0 $,
one obtains the condition
\begin{eqnarray}
\chi_0^{-1} (\omega \rightarrow 0) = - I_{\bf Q} .
\label{separatrix-pd}
\end{eqnarray}
In this paper we have carried out an asymptotic low-energy analysis
of our dynamical equations.
By contrast,
$\chi_0^{-1} (\omega \rightarrow 0)$
depends (through the Kramers-Kronig relation) on the solution at all energies.
Therefore, the determination of the phase boundary and, hence, the value of
the exponent $\alpha$, requires a complete (numerical) treatment of
the problem.
Recently, such a treatment has been carried out for a Kondo lattice model
with Ising RKKY couplings.\cite{Grempel:02,Zhu:03}
A locally critical quantum phase transition was identified,\cite{Grempel:02}
with a near-universal exponent $\alpha=0.72\pm 0.01$.
Extension of such numerical work to the isotropic case considered in the
present paper remains an important problem for the future.

Fig.~\ref{phase-diagram} also illustrates the important point that,
in a given system,
there is only one parameter that must be tuned to drive the system
through criticality.
The dashed line represents such a tuning trajectory.

Two additional remarks are in order.
First, the self-consistent
Weiss field $\chi_0^{-1}$, given in Eq.~(\ref{solution-2D-weiss}),
has a logarithmic dependence. In (imaginary) time, this corresponds to an
interaction with an algebraic component, $1/\tau$, multiplied by
a logarithmic correction.
On the other hand, the self-consistent solution for
the spin-spin correlation function at the critical point, given by
Eq.~(\ref{solution-2D-chi-loc}), corresponds to $1/\tau$, with no
logarithmic correction. Within the ($1\!-\!\gamma$)-expansion carried out
in Section~\ref{sec:impurity}, the logarithmic corrections to the range
of the interaction do not affect the critical exponent for
$\chi_{\text{loc}}$,
which remains algebraic without any logarithmic corrections.
We believe that this remains true beyond
the ($1\!-\!\gamma$)-expansion, since the spin-spin correlation functions
at a critical point are in general algebraic and admit
no logarithmic corrections.
For the particular form of the self-consistent Weiss-field
$\chi_0^{-1}$ used here, we can demonstrate this point explicitly
in a spherical model with a long-range interaction specified
by $\chi_0^{-1}$ (see Appendix~\ref{sec:appen-spherical}).

Second, the self-consistency equation~(\ref{self-consistent2D-2})
dictates that, at the critical point, the imaginary part of the
spin-self-energy $\text{Im}M(\omega)$ is sub-leading compared to both
$\text{Im}\chi_0 (\omega) $ and $\text{Im}\chi_{\text{loc}} (\omega)$.
In principle,
we should allow a sub-leading term in $\chi_0 (\omega)$, and determine
both the leading and the sub-leading terms for the critical
$\chi_{\text{loc}} (\omega)$. Calculating the sub-leading terms of
$\chi_{\text{loc}} (\omega)$ is, however, beyond the scope of the
($1\!-\!\gamma$)-expansion. What we have done, instead, is to take advantage
of the freedom to choose the Weiss field such that the self-consistency
equation is satisfied; Eq.~(\ref{self-consistent2D-2}), then,
allows us to determine $M$ from the leading term of
$\chi_{\text{loc}} (\omega)$.

\subsection{Self-consistent solution in three dimensions}
\label{sec:self-consistent3D}

If the RKKY density of states increases at the lower band edge
in a square-root fashion,
\begin{eqnarray}
\lim_{\epsilon \rightarrow I_{\bf Q}^+}
\rho_I (\epsilon ) \propto \sqrt{ \epsilon - I_{\bf Q}}
,
\label{rkky-dos-3d-edge}
\end{eqnarray}
the self-consistency equation~(\ref{self-consistent-2})
can again be written as in Eq.~(\ref{self-consistent-2D-sep}),
the difference being
that $\Lambda_0$ is now defined such that the square-root form
applies for $\epsilon \in (I_{\bf Q}, I_{\bf Q}+\Lambda_0)$.
Again, for a generic
$\rho_I(\epsilon)$,
$\Lambda_0$ is of the order of the RKKY interaction
$|I_{\bf Q}|$.

The second term on the right-hand side of Eq.~(\ref{self-consistent-2D-sep})
is finite, just as in two dimensions.
In three dimensions, however, the first term---when combined with
Eqs.~(\ref{M-QCP}) and~(\ref{rkky-dos-3d-edge})---also yields a finite value:
\begin{eqnarray}
\int_{I_{\bf Q}}^{I_{\bf Q}+\Lambda_0}
d \epsilon \frac {\rho_I (\epsilon) }
  { M(\omega) + \epsilon } \propto \sqrt{\Lambda_0} .
\label{self-consistent-3D-sep-1}
\end{eqnarray}

Thus, self-consistency dictates that in the three-dimensional case,
the local susceptibility $\chi_{\text{loc}}$ is finite at the QCP.
Hence, from Section~\ref{sec:cf}
[see the discussion concerning Fig.~\ref{impurity-qcp}(a)],
we know that the dynamics of the local moment are not yet critical.
Equivalently, the local energy scale $E_{\text{loc}}^*$ is finite at the QCP.

We now proceed to extract the critical behavior. To be
concrete,
we work with a semi-circular form for the RKKY density of states:
\begin{eqnarray}
\rho_I (\epsilon) = {2 \over \pi I^2} \sqrt{I^2 - \epsilon^2},
\label{semi-circle}
\end{eqnarray}
for $ | \epsilon| \le I$.
Solving the self-consistency equation~(\ref{self-consistent-2})
yields
\begin{eqnarray}
M(\omega) = \left ( { I \over 2} \right )^2
\chi_{\text{loc}} (\omega) + { 1 \over \chi_{\text{loc}}(\omega)},
\label{semi-circle-sc}
\end{eqnarray}
which, when combined with Eq.~(\ref{spin-se}), gives
\begin{eqnarray}
\chi_0^{-1} (\omega) = (I/2)^2 \chi_{\text{loc}}(\omega) .
\label{semi-circle-wiess}
\end{eqnarray}

The local physics is noncritical at the QCP, so
the imaginary part of the spin self-energy should reflect the decay of spin
fluctuations into particle-hole excitations and, hence, should be linear
in frequency. We can quantify $E_{\text{loc}}^*$ 
using the definition\cite{SiSmithIngersent}
\begin{eqnarray}
\text{Im} M^{-1} (\omega) = \omega / (E_{\text{loc}}^*)^2
\quad \mbox{for $|\omega| < E_{\text{loc}}^*$}.
\label{imM-3D}
\end{eqnarray}

Combined with Eq.~(\ref{M-QCP}), which now reads
$M(\omega\!\rightarrow\!0) = I$, this implies that
\begin{eqnarray}
M (\omega) = I - i (I/E_{\text{loc}}^*)^2 \omega + O ( \omega^2) .
\label{M-3D}
\end{eqnarray}
The precise value of $E_{\text{loc}}^*$ can be determined using a
technique that is applicable to the strong-coupling regime
of the Bose-Fermi Kondo Hamiltonian, Eq.~(\ref{H-imp});
one such
method is slave-boson mean-field theory.\cite{Hewson}

Combining Eqs.~(\ref{semi-circle-sc}) and~(\ref{M-3D}) leads to
the following form for the local susceptibility:
\begin{eqnarray}
\chi_{\text{loc}} (\omega + i0^+) =
&& {2 \over I} - {2 \over E_{\text{loc}}^*}\sqrt{2 \over I} \sqrt {-i\omega}
\nonumber\\
&&- 2 i \omega / ( E_{\text{loc}}^* )^2 + O (\omega^{3/2}) .
\label{chi-loc-3D}
\end{eqnarray}
Using Eq.~(\ref{semi-circle-wiess}), the Weiss field is
\begin{eqnarray}
\chi_0^{-1}(\omega+i0^+) =
&& {I \over 2 } - {2 \over E_{\text{loc}}^*} \left ({I \over 2} \right )^{3/2}
\sqrt {-i\omega} \nonumber\\
&&- 2 i (I / E_{\text{loc}}^* )^2  \omega + O (\omega^{3/2}) .
\label{chi-0-3D}
\end{eqnarray}

We should stress that Eqs.~(\ref{chi-loc-3D}) and~(\ref{chi-0-3D})
are compatible. In other words, for a given Bose-Fermi Kondo Hamiltonian
[Eq.~(\ref{H-imp})] and given bath dispersions
[Eqs.~(\ref{dos-fermion}) and~(\ref{dos-boson}) with $\gamma=1/2$],
the solution will yield an impurity dynamical spin susceptibility
whose imaginary part is proportional to
$|\omega|^{1/2}\text{sgn}\,\omega$. While the correlation functions on the
strong-coupling
side of the Bose-Fermi Kondo model are not accessible
within the $(1\!-\!\gamma)$-expansion (and are in general difficult to
determine except for certain large-N limits),
we note that the above is consistent with the bound set by
Griffiths' theorem.\cite{Griffiths}
The correlation functions in the paramagnetic
phase of the one-dimensional long-ranged spherical model
provide
additional support for this conclusion; see Appendix 
\ref{sec:appen-spherical}.

\section{Dynamical spin susceptibility of the locally critical point
at finite temperatures}
\label{sec:finite-T-lcp}

We now return to the locally critical point that occurs in the
two-dimensional case. Inserting our $T=0$ solution for the spin self-energy,
Eq.~(\ref{solution-2D-M}), into Eq.~(\ref{chi-edmft}), we arrive at
the following result for the momentum-dependent dynamical spin
susceptibility at $T=0$:
\begin{eqnarray}
\chi ({\bf q},\omega,T=0) = { 1 \over
{I_{\bf q} - I_{\bf Q} + \Lambda_0 (-i\omega / \Lambda)^{\alpha}}} ,
\label{chi-lcp-T0}
\end{eqnarray}
where $\Lambda_0$ is defined right after 
Eq.~(\ref{self-consistent-2D-sep}).
We emphasize once again that the local susceptibility,
given in Eq.~(\ref{solution-2D-chi-loc}),
is universal. On the other hand, the exponent $\alpha$ is
not universal, as discussed
after Eq.~(\ref{solution-2D-alpha}).
It should also be noted that, for ${\bf q \sim 0}$, 
Eq. (\ref{chi-lcp-T0}) is expected to be valid only if
the total spin is not conserved; in heavy fermions,
this is the case as a result of the strong spin-orbit
couplings. In the conserved case, it would be important
to keep track of the momentum-dependence of
the spin self-energy for small ${\bf q }$.

At finite temperatures, the self-consistent solution still has
$\gamma=0^+$. From Eq.~(\ref{chi}), we have
\begin{eqnarray}
\chi_{\text{loc}}(\tau) = {1 \over 4} { {\pi / \Lambda \beta } 
\over
{ \sin ( \pi \tau / \beta)} } ,
\label{chi-loc-tau-lcp}
\end{eqnarray}
yielding
\begin{eqnarray}
\chi_{\text{loc}}(\omega+i0^+,T)
={ 1 \over 2 \Lambda} \left[ \ln {\Lambda \over
{2\pi}T} - \psi \left({1 \over 2} - i {\omega \over {2\pi T} } \right) \right] ,
\nonumber\\
\label{chi-loc-omega-T-lcp}
\end{eqnarray}
where $\psi$ is the digamma function.

The corresponding spin self-energy can be found using
the second equality of Eq.~(\ref{self-consistent2D-2}).
The result is
\begin{eqnarray}
M(\omega+i0^+,T)
=&& - I_{\bf Q} +
\Lambda_0(2 \pi / \Lambda)^{\alpha}
T^{\alpha} {\cal M}(\omega/T) ,
\label{M-omega-T-lcp}
\end{eqnarray}
where
\begin{eqnarray}
{\cal M}(\omega/T)
= \exp [\alpha \psi (1/2 - i\omega / 2\pi T)] .
\label{cal-M-omega-T-lcp}
\end{eqnarray}

The momentum-dependent dynamical spin susceptibility at finite
temperatures
follows from inserting Eq.~(\ref{M-omega-T-lcp})
into Eq.~(\ref{chi-edmft}):
\begin{eqnarray}
\chi ({\bf q},\omega,T) = { 1 \over
{I_{\bf q} - I_{\bf Q} + \Lambda_0
(2 \pi / \Lambda)^{\alpha} T^{\alpha}
{\cal M}(\omega/T)}} .
\label{chi-q-omega-T-lcp}
\end{eqnarray}

\section{Ginzburg-Landau theory}
\label{sec:GL}

\subsection{Gaussian critical point and its instability}

In the picture of the $T=0$ SDW transition, there is only
one type of critical degrees of freedom, namely, the long-wavelength
fluctuations of the magnetic order parameter.
The Ginzburg-Landau (GL) action is an extension of the standard $\phi^4$
theory for a classical phase transition to allow for order-parameter
fluctuations that take place not only in space but also in (imaginary)
time.\cite{Hertz,Millis,Sachdev-book,Chubukov}
The temporal fluctuations can be thought of as adding $z$ dimensions,
where $z$ is the dynamic exponent.
In the antiferromagnetic case (${\bf Q}$ being nonzero),
$z=2$ and the action takes the form
\begin{eqnarray}
{\cal S}_{\text{SDW}} && =
{\cal S}_{\text{lw}} \,[\, {\bf m}
({\bf q} \sim {\bf Q}, \omega ) \,]
\nonumber \\
&& = \int d {\bf q} \int d \omega
\left[ r + ({\bf q} -{\bf Q})^2 + |\omega| \right]
[\, {\bf m} ({\bf q}, \omega ) \,]^2
\nonumber\\
&&+\, u \, \prod_{i=1}^{4} \int \!\! d {\bf q}_i \! \! \int \!\! d \omega _i
\, \delta\!\left(\sum_i {\bf q}_i\right) \delta\!\left(\sum_i \omega _i\right)
[\, {\bf m} \,]^4 \nonumber\\
&&+ \ldots
\label{GL-sdw}
\end{eqnarray}

The effective dimensionality is $d_{\text{eff}}=d+z$.
For $d=2$ or $d=3$, $d_{\text{eff}}$ equals or exceeds
the upper critical dimension of
$4$,
so the critical point is Gaussian.
The $T=0$ dynamical spin susceptibility assumes the mean-field
(RPA) form:
\begin{eqnarray}
\chi^{G}({\bf q}, \omega) \sim { 1 \over
{ ({\bf q} - {\bf Q})^2 - i \omega } },
\label{chi-sdw}
\end{eqnarray}
where the linear-in-frequency dependence comes from Landau damping
(decay into particle-hole pairs).
All the nonlinear couplings are irrelevant,
so the corresponding relaxation rate is superlinear, leading to a
violation of $\omega/T$ scaling.\cite{Sachdev-book,Millis,questions,d=z=2}
(This is very different from QCPs 
in insulators in two dimensions,\cite{Sachdev-Ye,Chakravarty}
where the dynamic exponent is usually $z=1$ and the corresponding
$\phi^4$-theory is below its upper critical dimension.)

It should be noted that for
two-dimensional magnetic fluctuations,
the local (i.e., $\bf q$-averaged) dynamical spin susceptibility
associated with $\chi^{G}({\bf q}, \omega)$ is, in fact, singular:
\begin{eqnarray}
\chi^{G}_{\rm \text{loc}}
(\omega )\equiv \int \!\! d {\bf q} \;\chi^{G} ({\bf q}, \omega)
\sim \ln { 1 \over {-i\omega}} .
\label{chi-sdw-local}
\end{eqnarray}
Within the SDW description [Eq.~(\ref{GL-sdw})], the nonlinear terms
couple only the long-wavelength modes, so
this singular $\chi^{G}_{loc}$ is inconsequential. 
However, when local degrees of freedom play
a central role from the outset---as
is the case for the local
moments in heavy fermions---this singularity can cause important
nonlinear effects in the dynamics of the local objects.
The preceding sections amount to a microscopic analysis of these nonlinear
effects.

\subsection{Ginzburg-Landau action for the locally critical point}

We now address
the important question of
whether the locally critical point is internally consistent
beyond the microscopic (EDMFT) approximations.

The defining feature of a locally critical point is the coexistence of
local and long-wavelength critical modes.
The long-wavelength modes, as usual, capture the divergence of
the correlation length in space. They reflect the fact that there is a
broken symmetry in the antiferromagnetic metal, and the order parameter
corresponding to this broken symmetry vanishes continuously as the critical
point is approached from the ordered side.
The local critical modes are additional critical modes. They arise from
a subtle ``ordering'' on the paramagnetic metal side, which
also disappears continuously as the critical point is approached
from the paramagnetic side. The ``ordering'' is associated with the
formation of Kondo singlets, a process which yields Kondo resonances and
makes the local moments part of the electron fluid. It is reflected in the
formation of a ``large'' Fermi surface, which encloses a volume
that counts both the number of conduction electrons and the
number of local moments.\cite{Hewson} The strength of these Kondo
resonances is captured by a quasiparticle residue $Z$,
associated with the states on such a large Fermi surface.
Unlike the SDW case, at the locally critical point $Z$ goes to zero
over the entire large Fermi surface.

The GL action, then, has to be the sum of three parts:
\begin{eqnarray}
{\cal S}_{\text{LCP}} =
{\cal S}_{\text{lw}}\,[\, {\bf m} ({\bf q} \sim {\bf Q}, \omega ) \,]
+ {\cal S}_{\text{loc}}
+ {\cal S}_{\text{mix}} .
\label{GL-lcp}
\end{eqnarray}
Here, ${\cal S}_{\rm lw}$ is the same as in Eq.~(\ref{GL-sdw}),
${\cal S}_{\text{loc}}$ describes
the couplings among the local critical modes alone,
and ${\cal S}_{\text{mix}}$
takes into account the coupling between the local
and spatially extended degrees of freedom.

The notion that local fluctuations are part of the quantum
critical theory is at first sight surprising,
especially viewed from what is known about classical
(finite-temperature) phase transitions. 
The universality class of a classical critical point is
entirely determined by statics.\cite{Goldenfeld}
(Real-time dynamics are considered only after the statics have
determined the universality class.\cite{HH})
Static fluctuations can be energetically favorable only if they occur
at long distances.
The universality class of a 
QCP,
by contrast,
depends on statics and dynamics (quantum fluctuations)
on an equal footing. The mixing of statics and dynamics opens
the door for local fluctuations to have low energies, provided
that their fluctuations in (imaginary) time are slow.

We are now in a position to address the robustness of the locally
critical point.
In the microscopic analysis, we assumed that the spin self-energy
and the electron self-energy are independent of the wavevectors $\bf q$
and $\bf k$.
Allowing a $\bf k$ dependence in the conduction-electron
self-energy will not affect our analysis, since we are away from
half-filling and the conduction-electron density of states is expected
to remain finite. We now argue that allowing a $\bf q$ dependence
in the spin self-energy leaves the locally critical point
intact, provided that $\alpha < 1$ (as is the case observed
in CeCu$_{6-x}$Au$_x$ and YbRh$_2$Si$_2$).
Consider first the contribution to the spin self-energy
from the nonlinear couplings within ${\cal S}_{\rm lw}$.
The effective dimension $d_{\text{eff}}=d+z=2+2/\alpha$ still
exceeds $4$, so all the nonlinear couplings within $S_{\rm lw}$
must be irrelevant in the RG sense.
The $\bf q$ dependence of the spin self-energy
will then be at most $({\bf q}- {\bf Q})^2$.
Consider next the contribution to the spin self-energy from
${\cal S}_{\text{mix}}$.
Since the spin modes are coupled to the local modes, the corresponding
contribution cannot have a singular $\bf q$ dependence, either.
As a result, the spatial anomalous dimension $\eta$ takes the value
$\eta = 0$, and
the zero-temperature, zero-frequency spin susceptibility
for ${\bf q} \sim {\bf Q}$ goes as
\begin{eqnarray}
\chi({\bf q}) \sim { 1 \over
{ ({\bf q} - {\bf Q})^2 }} .
\label{chi-q}
\end{eqnarray}
The corresponding local susceptibility remains singular.
Therefore, the local criticality is robust.

We emphasize that the locally critical point is non-Gaussian because the
appropriate field theory is {\it not} just the usual $\phi^4$ theory, but
instead has the form given in Eq.~(\ref{GL-lcp}) above.
Each of the three parts making up ${\cal S}_{\text{LCP}}$ contains
nonlinear couplings which flow under the RG transformation.
${\cal S}_{\text{lw}}$ (the $\phi^4$ theory) contains nonlinear couplings
purely between long-wavelength modes.
Since the effective dimensionality satisfies $d_{\text{eff}}=d+z > 4$,
all these couplings are irrelevant, and they will contribute to the spin
relaxation rate a term that is superlinear in $T$.\cite{Millis,Sachdev-book}
However, ${\cal S}_{\text{loc}}$ and ${\cal S}_{\text{mix}}$ contain
nonlinear couplings involving the local modes.
Some of these couplings are relevant, and will generate a linear-in-$T$
contribution to the spin relaxation rate. The total relaxation rate will be
dominated by this second type of contribution, and will be linear in $T$,
giving rise to $\omega/T$ scaling in the dynamical spin susceptibility.
Our microscopic theory
provides
a specific prescription for determining
this relaxation rate, as well as the universal scaling functions.

We close this subsection with a brief comment on the general validity
of the EDMFT analysis of quantum critical behavior in metals.
We have seen that, for the EDMFT results to be valid,
it is crucial\cite{Nature} that the critical point has
a mean-field exponent for spatial fluctuations, i.e., $\eta=0$.
(The temporal fluctuations are allowed to have an anomalous exponent.)
This requires the nonlinear couplings among the long-wavelength
modes to be irrelevant, a condition that is relatively easy to satisfy
at
a $T=0$ transition in a metallic system.
The results presented above serve as a concrete example where the spatial
fluctuations are Gaussian but the temporal fluctuations are non-Gaussian.
For critical points where $\eta \ne 0$, by contrast, the EDMFT
is not expected to describe the correct critical properties.
For instance, at finite-temperature transitions (where
$\eta$ is in general nonzero), the EDMFT yields a first-order
transition.\cite{Chitra-Si,Pankov,Zhu:03,Sun,Burdin}
However, quantum critical dynamics---including those at finite
temperatures---are controlled by the QCP at $T=0$, and are correctly
captured by EDMFT calculations.

\section{Comparison with experiments}
\label{sec:expt}

We now compare our conclusions with
existing
experiments in heavy-fermion
metals, and also make several predictions that can be tested
in the future.

\subsection{Inelastic neutron scattering, magnetization,
and NMR relaxation rate}

Our theoretical result for the dynamical spin susceptibility near a
locally critical point [Eq.~(\ref{chi-q-omega-T-lcp})] can be rewritten
\begin{eqnarray}
\chi ({\bf q},\omega,T) = { 1 \over
{ 
f({\bf q})
+ A T^{\alpha}
{\cal M}(\omega/T) }} ,
\label{chi-q-omega-T-lcp2}
\end{eqnarray}
where $A = \Lambda_0 (2 \pi / \Lambda)^{\alpha}$,
and $f({\bf q}) = I_{\bf q} - I_{\bf Q}$ vanishes 
at the peak wavevector (${\bf q}= {\bf Q}$) but is
finite at other wavevectors.
Equation~(\ref{chi-q-omega-T-lcp2}) implies that the
dynamical spin susceptibility at the peak wavevector
satisfies an $\omega/T$ scaling:
\begin{equation}
 \chi({\bf Q}, \omega, T) = {1 \over A T^{\alpha}
 {\cal M}(\omega/T)} ,
\label{w/T}
\end{equation}
where the scaling function ${\cal M}(\omega/T) $
is given in Eq.~(\ref{cal-M-omega-T-lcp}).
Eq.~(\ref{w/T}) is consistent with the 
inelastic neutron scattering
results\cite{Schroder2,Schroder1}
on CeCu$_{6-x}$Au$_x$ at the critical
concentration $x_c \approx 0.1$, where it is
found that
the exponent $\alpha \approx 0.75$.
($\omega/T$ scaling was reported\cite{Aronson} 
earlier in UCu$_{5-x}$Pd$_x$ for $x=1$ and 1.5.
This system, however, seems to contain strong disorder, the effects of which
remain a subject of debate.\cite{Aronson,MacLaughlin,Miranda,Neto})
The scaling function used
to fit the experiments
is 
$(1-i\omega/aT)^{\alpha}$, which behaves very similarly to 
Eq.~(\ref{cal-M-omega-T-lcp}).
The neutron scattering results\cite{Schroder2,Schroder1}
at generic wavevectors have
also been fitted to the form of Eq.~(\ref{chi-q-omega-T-lcp2}),
with the same
exponent $\alpha\approx 0.75$.

In our theory, the locally critical picture arises if the magnetic
fluctuations are two-dimensional.
Within experimental resolution, 
$f({\bf q})$ goes to zero along lines in the three-dimensional
Brillouin zone, implying that the magnetic fluctuations
in the quantum critical regime are two-dimensional in real space
in the frequency and temperature ranges that have been studied
to date.\cite{Stockert,Stockert2,Schroder2}
It should be noted, however, that the magnetic ordering for $x> x_c$ is
three-dimensional.
To properly address the finite-$T$ magnetic ordering
transition, it will be necessary to incorporate the (small)
RKKY coupling in the third dimension.\cite{note-dimensionality}
Whether dimensional crossover eventually takes place 
at lower temperatures in the quantum critical regime
is a question that should be addressed in future experiments.
We hope that our work will provide
a stimulus for this kind of (challenging) experiment.
It is also desirable that neutron scattering be carried out at a QCP
where the magnetic fluctuations are genuinely three-dimensional.
In this regard, we note that 
a violation of $\omega/T$ scaling has been reported\cite{Flouquet}
in Ce$_{1-x}$La$_x$Ru$_2$Si$_2$.

It also follows from Eq.~(\ref{chi-q-omega-T-lcp2}) that the 
static and uniform spin susceptibility, $\chi \equiv M/H$,
has a modified Curie-Weiss form:
\begin{equation}
 \chi(T) = {1 \over {\Theta + B \, T^{\alpha}}} ,
\label{chi(T)}
\end{equation}
where $\Theta = I_{{\bf q}={\bf 0}} - I_{\bf Q}$ and
$B \approx \Lambda_0\Lambda^{-\alpha}
(2 \pi)^{\alpha}$ $\exp[\alpha \psi(1/2) ]$. 
Eq.~(\ref{chi(T)}) fits well\cite{Schroder2,Schroder1} 
the magnetization data in CeCu$_{6-x}$Au$_x$.

There are indications that the modified Curie-Weiss susceptibility describes
several other heavy-fermion metals near a QCP.\cite{StewartRMP}
One such material is YbRh$_2$Si$_2$,\cite{Gegenwart}
which exhibits thermodynamic and transport properties very similar
to those of CeCu$_{6-x}$Au$_x$, and has a layered structure that suggests
the possibility of (quasi-)two-dimensional magnetic fluctuations.
These characteristics make YbRh$_2$Si$_2$ a promising candidate for local
criticality.
However, inelastic neutron scattering has yet to be performed in this system.

The local susceptibility given in Eq.~(\ref{chi-loc-omega-T-lcp})
leads to a non-Korringa NMR spin-lattice relaxation rate,
\begin{equation}
{1 \over T_1} \sim {\rm constant}.
\label{T1}
\end{equation}
(We have assumed that the hyperfine constant is not strongly
${\bf q}$-dependent.)
This prediction might be tested
using NMR measurements on the Cu sites in CeCu$_{6-x}$Au$_x$ or
the Si sites in YbRh$_2$Si$_2$.

\subsection{Fermi surface properties, Hall coefficient and other transport
properties}

On the paramagnetic side
of a magnetic ordering QCP,
the coherence energy scale ($E_{\text{loc}}^*$)
is finite. This signals\cite{Hewson} the formation of heavy quasiparticles,
whose Fermi surface is ``large'' in the sense that its volume counts both
the conduction electrons and the local moments.
In the case of an SDW transition, $E_{\text{loc}}^*$ remains finite 
through the QCP, and the Fermi surface remains large on the 
antiferromagnetic side (except for the folding of the Fermi surface due
to a broken translational symmetry). 
By contrast, $E_{\text{loc}}^*$ vanishes at a locally critical point,
and the Fermi surface is ``small'' on the antiferromagnetic
side,\cite{E-loc-ordered}
i.e., its volume and topology are such that the
single-electron excitations come entirely from
the conduction electrons.
Thus, the topology of the Fermi surface differentiates
the antiferromagnetic metal phase on the ordered side 
of the locally critical transition from
its counterpart for an SDW transition.
Experimental probes of the Fermi surface, such as de Haas-van Alphen
measurements, both inside the antiferromagnetic metal phase and close
to the QCP, may be used to distinguish
between the two types of quantum phase transition.

As a consequence of the large change in the Fermi-surface
volume across a locally critical point, the Hall coefficient 
at zero temperature is expected to undergo a large jump at the QCP.
This provides another potential test of the locally critical picture.
(A more general discussion of the behavior of the Hall coefficient
can be found in Ref.~\onlinecite{questions}; 
see also Ref.~\onlinecite{Yeh}.)

Finally, we consider the longitudinal resistivity.
At an antiferromagnetic SDW critical point, only
certain
hot spots
of the Fermi surface are affected by the critical fluctuations.
In the clean limit, then, the resistivity is expected to retain
a Fermi-liquid $T^2$ form.\cite{Hlubina,Rosch} At a locally critical
point, on the other hand, the entire Fermi surface is affected and
the resistivity will have a non-Fermi-liquid form.

The change of the Fermi-surface volume is also expected to be
manifested in the behavior of the residual resistivity as a function
of the tuning parameter. This is best studied by using
pressure or a magnetic field as the tuning parameter.
(By contrast, tuning by chemical doping would also induce a significant change
in the amount of disorder.)
We note that the residual resistivity does seem to show dramatic changes
across the pressure-driven QCP in CeCu$_{\rm 5}$Au 
(Ref.~\onlinecite{Wilhelm}) 
and the field-driven QCP in YbRh$_{\rm 2}$Si$_{\rm 2}$ 
(Ref.~\onlinecite{rho0-ybrh2si2}).

\section{Conclusions and outlook}
\label{sec:conclusion}

In this paper, we have identified a new class of quantum critical point
in Kondo lattice systems. We have carried out
an EDMFT analysis of the Kondo lattice model and found that,
when the magnetic fluctuations are two-dimensional, a locally
critical point arises. Here, critical local modes coexist
with the long-wavelength fluctuations of the order parameter.
We have also argued that the locally critical point is robust beyond 
our microscopic calculations.

Our results explain the salient features of the experiments in
several heavy-fermion systems. In particular,
we have made comparisons with the existing
experiments in CeCu$_{6-x}$Au$_x$ and YbRh$_2$Si$_2$. We have 
also made a number of predictions concerning the NMR relaxation
rate, the Fermi-surface volume, and the Hall coefficient.

The crucial ingredient of local criticality is
the presence of spatially local critical modes.
This contrasts with the standard critical theory, in which the
only low-energy modes are long-wavelength fluctuations
of the order parameter. For heavy fermions,
the emergence of such critical local fluctuations
can ultimately be traced to
the fact that, as a result of
the
microscopic Coulomb blockade,
local moments are ``pre-formed'' at intermediate energies.\cite{Anderson}
At an SDW transition, the local moments disappear at sufficiently
low energies; their quenching by the conduction-electron spins leaves
only electron-like excitations (Kondo resonances).
At a locally critical point, on the other hand, vestiges of the
local moments persist to asymptotically low energies. Such local-moment
physics is neglected in the SDW treatment of magnetic quantum critical
points. 

For Mott-Hubbard systems, the microscopic Coulomb blockade
is responsible for the formation of the Mott insulator.
One
is therefore led to
speculate that low-energy local modes may well play
a key role in metals near a Mott transition.
This issue is becoming increasingly important due to the 
observations of (nearly) quantum critical behavior in transition-metal
oxides,\cite{Mackenzie} as well as the apparent scaling behavior
in high-temperature superconductors.\cite{VarmaNFL,Tallon,Johnson,Aeppli}

\acknowledgments

We thank E.\ Fradkin and S.\ Sachdev for useful discussions
concerning logarithmic corrections in the Bose-Fermi Kondo model.
This work has been supported in part
by NSF Grant No.\ DMR-0090071 and TCSUH. Q.S.\ would also like to
acknowledge the support of Argonne National Laboratory, the University
of Chicago, and the University of Illinois at Urbana-Champaign during
his sabbatical leave, and the hospitality of the 
Aspen Center for Physics.

\vskip 0.2 in
\appendix

\section{Derivation of the RG equations}
\label{sec:appen-rg}

In this appendix, we present the derivation of the RG equations~(\ref{rg-eq}),
following a procedure outlined in Ref.~\onlinecite{Smith3}.

Consider first the pseudo-$f$-electron self-energy.
There are two contributions,
as specified by Fig.~\ref{f-self-energy}. The contribution~(a) is given by
\begin{eqnarray}
\Sigma^{(a)}(i\omega_f) =
&&-{3 \over 8} {J_K^2 \over \beta^2} \sum_{p,p'}
\sum_{i\omega_1} \sum_{i\omega_2}
 G^b_f(i\omega_1) \times \nonumber\\
&&\times G^b(p,i\omega_2)G^b(p',i\omega_f+i\omega_2-i\omega_1) .
\label{sigma-a}
\end{eqnarray}
Here the factor of 3 comes from summation over the three spin components.
Carrying through the summation over $i\omega_1$ and $i\omega_2$,
setting $i\omega_f = \omega + \lambda$, and using the fact that
$\lambda \rightarrow \infty$, we have
\begin{eqnarray}
\Sigma^{(a)}(\omega + \lambda ) && =
{3 \over 8} J_K^2 \sum_{p,p'}
{{f(\epsilon_p) [1-f(\epsilon_{p'})]}
\over {\omega + \epsilon_p - \epsilon_{p'}}} \nonumber\\
&&=
-{3 \over 8} (N_0J_K)^2  \! \left[ 2 D \ln 2 + \omega \ln {D \over \omega}
+ O\!({\omega})\right],
\label{sigma-a1}
\end{eqnarray}
where $D$ is the half-bandwidth,
and the Fermi factor $f(\epsilon ) = ( e^{\beta\epsilon} + 1)^{-1}$.
Note that in the second line we have set $T=0$, since our
purpose here is to construct the RG equations.

Similarly,
\begin{eqnarray}
\Sigma^{(b)}(i\omega_f) =
-{3 \over 4} {g^2 \over \beta} \sum_{p}
\sum_{i\omega_1}  G^b_f(i\omega_1) G^b_{\phi}
(p,i\omega - i\omega_1) .
\label{sigma-b}
\end{eqnarray}
To linear order in $1-\gamma$, we can set $\gamma=1$ in
$G^b_{\phi}$ and the corresponding spectral function is
\begin{eqnarray}
A_{\phi}(\epsilon) &\equiv&
-{1 \over \pi}\text{Im} \sum_p G^b_{\phi}(p, \epsilon+i0^+) \nonumber \\
&=& K_0^2 \epsilon  \quad
\mbox{for $|\epsilon|<\Lambda$}.
\label{A-phi}
\end{eqnarray}
We then have
\begin{eqnarray}
\Sigma^{(b)}(\omega+\lambda) && =
{3 \over 4} g^2 \int d\epsilon
A_{\phi}(\epsilon) 
{{1+n_B(\epsilon ) }
\over {\omega - \epsilon}} \nonumber\\
&&= -{3 \over 4} (K_0g)^2 
\left[ \Lambda + \omega  \ln {\Lambda \over \omega}
+ O({\omega})\right ] ,
\label{sigma-b1}
\end{eqnarray}
where the Bose factor $n_B(\epsilon ) = (e^{\beta\epsilon} - 1)^{-1}$.
The constant parts can be absorbed in a shift
of
$\lambda$. The
$\omega \ln \omega$ parts are retained in Eq.~(\ref{Sigma-f}), where
we have replaced the cutoffs by the running cutoff $W$.

We now turn to the corrections to the Kondo coupling, as given in
Fig.~\ref{J-vertex}.
We set the external frequencies of the conduction
electrons to $0$, and those for the pseudo-$f$-electrons to $i\omega_f$.
The contribution from Fig.~\ref{J-vertex}(a) is
\begin{eqnarray}
\Gamma_J^{(a)}(\omega\!+\!\lambda) && =
- {J_K^2 \over 2 \beta} \sum_p
\sum_{i\omega_1}  G^b(p,i\omega_1) G^b_f(i\omega_f\!+\!i\omega_1)
|_{i\omega_f=\omega\!+\!\lambda} \nonumber\\
&&=-{1 \over 2} J_K^2 N_0 \int d \epsilon
{ { f(\epsilon) }
\over { \omega + \epsilon }}
= {1 \over 2} J_K^2 N_0 \ln { D \over \omega} ,
\label{gamma-J-a}
\end{eqnarray}
where, in the second equality (and at similar places in the equations
to follow) we take $\lambda \rightarrow \infty$.
The contribution from Fig.~\ref{J-vertex}(b) is
\begin{eqnarray}
\Gamma_J^{(b)}(\omega\!+\!\lambda) && =
{J_K^2 \over {2\beta}}
\sum_{i\omega_1}
\sum_p
G^b(p,i\omega_1) G^b_f(i\omega_f\!-\!i\omega_1)
|_{i\omega_f=\omega+\lambda} \nonumber\\
&&=-{1 \over 2} J_K^2 N_0 \! \int \!\! d \epsilon
{ { 1\!-\!f(\epsilon) }
\over { \omega\!-\!\epsilon  }}
= {1 \over 2} J_K^2 N_0 \ln { D \over \omega} .
\label{gamma-J-b}
\end{eqnarray}
The contribution from Fig.~\ref{J-vertex}(c) is
\begin{eqnarray}
\Gamma_J^{(c)}(i\omega_f) && =
-{J_K^3 \over 8 \beta}
\sum_{i\omega_1}  \chi_c (i\omega_f - i\omega_1) [G^b_f(i\omega_1)]^2 ,
\label{gamma-J-c}
\end{eqnarray}
where $\chi_c(i\omega)$ is the bare local conduction-electron
susceptibility, which has a spectral function
\begin{eqnarray}
A_{\chi_c} \equiv {1 \over \pi } \chi_c''(\epsilon+i0^+) = N_0^2 \epsilon .
\label{A-chi}
\end{eqnarray}
The result is
\begin{eqnarray}
\Gamma_J^{(c)}(\omega + \lambda)
&& = -{1 \over 8} J_K^3 \int d \epsilon A_{\chi_c}(\epsilon)
{{1 + n_B(\epsilon) } \over { (\omega - \epsilon )^2}}
\nonumber\\
&& =
-{1 \over 8} J_K^3 N_0^2 \ln { D \over \omega} .
\label{gamma-J-c1}
\end{eqnarray}
The contribution from Fig.~\ref{J-vertex}(d) is
\begin{eqnarray}
\Gamma_J^{(d)}(i\omega_f) && =
{{g^2J_K} \over {4\beta}}
\sum_{i\omega_1}\! \sum_p \! G^b_{\phi} (p, i\omega_f\!-\!i\omega_1)
[G^b_f(i\omega_1)]^2 .
\label{gamma-J-d}
\end{eqnarray}
Using Eq.~(\ref{A-phi}), we have
\begin{eqnarray}
\Gamma_J^{(d)}(\omega + \lambda)
&& = -{1 \over 4} g^2 J_K \int d \epsilon A_{\phi}(\epsilon)
{{1 + n_B(\epsilon) } \over { (\omega - \epsilon )^2}}
\nonumber\\
&& =
-{1 \over 4} J_K (K_0g)^2 \ln { \Lambda \over \omega} .
\label{gamma-J-d1}
\end{eqnarray}

Finally, we consider the corrections to the coupling constant $g$.
We set the external frequency for the $\phi$ propagator to $0$,
and that for each pseudo-$f$-electron propagator to $i\omega_f$.
Then
\begin{eqnarray}
\Gamma_g^{(a)}(\omega\!+\!\lambda) && =
-  {{g J_K^2} \over { 8\beta}}
\sum_{i\omega_1}  \chi_c (i\omega_f\!-\!i\omega_1)
[G^b_f(i\omega_1)]^2 |_{i\omega_f=\omega+\lambda} \nonumber\\
&&=- {1 \over 8} g (N_0J_K)^2 \ln { D \over \omega} ,
\label{gamma-g-a}
\end{eqnarray}
and
\begin{eqnarray}
\Gamma_g^{(b)}(\omega\!+\!\lambda) && =
 {g^3 \over 4 \beta}
\sum_{i\omega_1}  G^b_{\phi} (i\omega_f\!-\!i\omega_1)
[G^b_f(i\omega_1)]^2 |_{i\omega_f=\omega+\lambda} \nonumber\\
&&=- {1 \over 4} g (K_0g)^2 \ln { D \over \omega} .
\label{gamma-g-b}
\end{eqnarray}

Setting $\omega=W'$ in Eq.~(\ref{z}) leads to
\begin{eqnarray}
z_f &=& 1 + \left[ {3 \over 8}(N_0J_K)^2 + { 3 \over 4} (K_0g)^2 \right]
\ln{W \over W'}
\nonumber\\
z_J &=& 1 + \left[ N_0J_K - {1 \over 8}(N_0J_K)^2 - {1 \over 4} (K_0g)^2 \right]
\ln{W \over W'}
\label{z-fJg}
\\
z_g &=& 1 - \left[ {1 \over 8} (N_0J_K)^2 + {1 \over 4}(K_0g)^2 \right]
\ln{W \over W'}\nonumber
\end{eqnarray}
From Eq.~(\ref{j'}), we have
\begin{eqnarray}
J_K' &=& J_K + J_K \left[ N_0 J_K - { 1 \over 2}(N_0J_K)^2 - (K_0g)^2 \right]
\ln{W \over W'}\nonumber\\
g' &=&  g - g \left[ { 1 \over 2}(N_0J_K)^2 + (K_0g)^2 \right]
\ln {W \over W'}
\label{J'Jg'g}
\end{eqnarray}

\vskip 0.2 in

\section{Determination of the separatrix in the 
$J_K$-$\lowercase{g}$
plane from the RG flow}
\label{sec:appen-separatrix}

Our starting point is the RG equations~(\ref{rg-eq}),
which yield the RG flow shown in Fig.~\ref{rg-flow}.
We imagine approaching the separatrix
from the local-moment side. The separatrix is identified when the initial
parameters are such that $N_0 J_K$ no longer flows to zero but instead
is renormalized to its fixed-point value $\sqrt{(1-\gamma)/2}$.

Consider the separatrix near the origin. Provided that we work
either on the local-moment side or on the separatrix itself,
$(N_0J_K)^2$ is at most of the order of $(1-\gamma)^2$ during
the entire flow and can safely be ignored inside the brackets
on the right-hand sides of both flow equations~(\ref{rg-eq}).
(This is not the case on the Kondo side.)
With this approximation, and the shorthand notation
\begin{eqnarray}
a \equiv  N_0J_K, \quad
b \equiv  (K_0g)^2,
\label{ab}
\end{eqnarray}
the RG equations simplify to
\begin{eqnarray}
{{ d a } \over { d l }}
&& = a ( a - b ) ,
\label{rg-eq2.1}
\\[-1ex]
\nonumber
\\[-1ex]
{{ d b} \over  {d l }}
 && = 2 b \left[ {{1 - \gamma} \over 2} - b \right]
.
\label{rg-eq2.2}
\end{eqnarray}
These equations have an unstable fixed point at
\begin{eqnarray}
a^* = b^* = {{1 - \gamma} \over 2} .
\label{a*b*}
\end{eqnarray}

Solving Eq.~(\ref{rg-eq2.2}) yields
\begin{eqnarray}
b = {{1 - \gamma} \over 2} \left[
1 + ( c_0 -1 ) e^{-(l-l_0)(1-\gamma)} \right]^{-1} .
\label{b}
\end{eqnarray}
Here,
\begin{eqnarray}
c_0 \equiv { {1 - \gamma} \over {2 b_0}} ,
\label{c0}
\end{eqnarray}
and $b_0$ and $l_0$ (as well as $a_0$, $z_0$, and $y_0$ used below)
are the initial values of the corresponding parameters.

We can now proceed to solve Eq.~(\ref{rg-eq2.1}). Let us write
\begin{eqnarray}
a = z y,
\label{a=zy}
\end{eqnarray}
where the factor $z$ is defined as satisfying
\begin{eqnarray}
{{d z} \over {d l}} = - b z.
\label{z-def}
\end{eqnarray}
When combined with Eq.~(\ref{b}), Eq.~(\ref{z-def}) can be easily solved
to give
\begin{eqnarray}
z = z_0 \left[
{ { c_0 e ^{-(l-l_0)(1-\gamma)}}
\over
{1 + (c_0 -1 ) e ^{-(l-l_0)(1-\gamma)}}
}
\right]^{{1 \over 2}} .
\label{z-solution}
\end{eqnarray}
The other factor, $y$, satisfies the
differential equation
\begin{eqnarray}
{{d y} \over {d l}} = z y^2,
\label{y}
\end{eqnarray}
which, when combined with Eq.~(\ref{z-solution}), leads to
\begin{eqnarray}
y = y_0 \left[ 1 - {a_0 \over {1-\gamma} }
\sqrt{c_0} A(l) \right] ^{-1} ,
\label{y2}
\end{eqnarray}
where
\begin{eqnarray}
A(l) \equiv \int_1^{ e^{(l-l_0)(1-\gamma)} }
\!\!\!\! { d\lambda \over { \lambda \sqrt{\lambda + c_0 - 1} }} .
\label{A}
\end{eqnarray}
Using\cite{GandR}
\begin{eqnarray}
\int { dx \over
{\sqrt{x} (a+bx)}}
= { 1 \over {i\sqrt{ab}}}
\ln { {a - b x + 2 i \sqrt{abx} } \over {a + bx}}\nonumber\\
~~~~~~~~~~~~~~~~~~~~~~~~~~~~\text{for} ~ab < 0,
\label{GR}
\end{eqnarray}
and recognizing that, for parameters near the origin,
$c_0 \equiv (1-\gamma)/(2b_0) \ll 1$, we obtain
the following asymptotic form in the limit
$l-l_0 \gg 1/(1-\gamma)$:
\begin{eqnarray}
A(l\rightarrow \infty) \approx { 1 \over \sqrt{c_0} }
\ln (4c_0) - 2 e^{-(l-l_0)(1-\gamma)/2} .
\label{A-infty}
\end{eqnarray}
Combining Eqs.~(\ref{a=zy}), (\ref{z-solution}), (\ref{y2}),
and~(\ref{A-infty}),
we arrive at the asymptotic solution for $a$:
\begin{eqnarray}
a(l\rightarrow \infty) \approx 
{ { a_0 \sqrt{c_0} e^{-(l-l_0)(1-\gamma)/2 } }
\over
{B + {{ a_0 \sqrt{c_0} } \over a^* }
e^ {-(l-l_0)(1-\gamma)/2 } }},
\label{a-solution}
\end{eqnarray}
where
\begin{eqnarray}
B = 1-{a_0 \over {1-\gamma}} \ln (4c_0) .
\label{B}
\end{eqnarray}

It is now clear that when $B$ is positive, $a(l\!=\!\infty)=0$.
In other words, the system falls in the local-moment regime.
On the other hand, $a(l\!=\!\infty)=a^* $ when $B=0$, i.e., when
\begin{eqnarray}
a_0 \ln (4c_0) = 1 - \gamma.
\label{separatrix-condition}
\end{eqnarray}
This condition specifies the separatrix. Written in terms of the
original variables, through Eqs.~(\ref{ab}) and~(\ref{c0}), this
becomes Eq.~(\ref{separatrix}).

For completeness, we also write down the equation for the separatrix
when the condition $c_0 \gg 1$ is not satisfied. In this case, $a(l)$
still has the asymptotic form specified by Eq.~(\ref{a-solution}),
but Eq.~(\ref{B}) is replaced by
\begin{eqnarray}
B = 1-{ {a_0 \sqrt{c_0} } \over
{(1\!-\!\gamma ) \sqrt{c_0\! -\!1} }}
\ln \left[ 2c_0\! -\! 1\! +\! 2 \sqrt{c_0(c_0\! -\! 1)} \right],
\label{B-exact}
\end{eqnarray}
and Eq.~(\ref{separatrix-condition}) is modified accordingly.
It is straightforward to check that the choice $(a_0,b_0)=(a^*,b^*)$
satisfies the condition $B=0$. In other words, the unstable
fixed point is part of the separatrix, as it should be.

\vskip 0.2 in

\section{Spin-spin correlation function at the critical point:
$(1\!-\!\gamma)$-expansion}
\label{sec:appen-corr}

Here we calculation the spin-spin correlation function
\begin{eqnarray}
\chi_{\text{loc}} (\tau) \equiv { 1 \over 2}
\langle T_{\tau} S^{-} (\tau) S^{+}( 0 ) \rangle
= \lim_{\lambda \rightarrow \infty}
{1 \over 2} e^{\beta \lambda} \tilde{\chi} ,
\label{chi-loc-1}
\end{eqnarray}
where
$\tilde{\chi}$ is calculated in the pseudo-fermion representation:
\begin{eqnarray}
\tilde{\chi} (\tau) = { 1 \over 2}
\langle T_{\tau} f_{\downarrow}^{\dagger} f_{\uparrow}(\tau)
f_{\uparrow}^{\dagger} f_{\downarrow}(0)\rangle .
\label{chi-loc-2}
\end{eqnarray}
It turns out to be convenient to calculate $\tilde{\chi}$ in terms of
\begin{eqnarray}
G^b_f (\tau) = e^{-\lambda \tau}
[(f(\lambda)-1)\Theta(\tau) +
f(\lambda) \Theta(-\tau) ] .
\label{g-f-tau}
\end{eqnarray}

The contributions to $\tilde{\chi}$ are given in Fig.~\ref{chi-diagrams}.
Note that any diagram
containing more than one isolated $f$-electron bubble vanishes
once the limit $\lambda \rightarrow \infty$ is taken.
Fig.~\ref{chi-diagrams}(a) yields
\begin{eqnarray}
\tilde{\chi}^{(a)} (\tau) = - { 1 \over 2} G^b_f(\tau) G^b_f(-\tau) .
\label{chi-tilde-a}
\end{eqnarray}
In the following, we will restrict $0<\tau<\beta$. Then
\begin{eqnarray}
\tilde{\chi}^{(a)} (\tau) = { 1 \over 2} e^{-\beta \lambda} .
\label{chi-tilde-a1}
\end{eqnarray}
The contribution from Fig.~\ref{chi-diagrams}(b) is
\begin{eqnarray}
\tilde{\chi}^{(b)} (\tau) =
&&{ {  g^2 } \over 8}
\int_0^{\beta}d\tau_1 \int_0^{\beta} d \tau_2
G^b_{\phi}(\tau_1 + \tau_2) \times \nonumber\\
&& \times G^b_f(\tau_1) G^b_f(\tau_2) G^b_f(\!-\tau_1\!-\!\tau)
G^b_f(\!-\tau_2\!+\!\tau),
\label{chi-tilde-b}
\end{eqnarray}
and those from Figs.~\ref{chi-diagrams}(c) and \ref{chi-diagrams}(d) sum to
\begin{eqnarray}
\tilde{\chi}^{(cd)} (\tau) =
&&
- { { 3 g^2 } \over 8}
\int_0^{\beta}d\tau_1 \int_0^{\beta} d \tau_2
G^b_{\phi}(\tau-\tau_1 - \tau_2) \times \nonumber\\
&& \times G^b_f(\tau_1) G^b_f(\tau_2) G^b_f(\tau-\tau_1-\tau_2)
G^b_f(-\tau)
\label{chi-tilde-cd}
\end{eqnarray}
Anticipating that $g^2$ is of order $1-\gamma$, to linear order in
$1-\gamma$ we can use 
Eq.~(\ref{A-phi}),
which corresponds to
\begin{eqnarray}
G^b_{\phi} (\tau) 
&&= \int d \epsilon A_{\phi} (\epsilon) {1 \over \beta}
\sum_{i\omega_n} { 1 \over {i\omega_n - \epsilon} } 
{\rm e}^{-i\omega_n \tau} \nonumber \\
&& = K_0^2 \left(
{\pi/\beta}
\over {\sin (\pi \tau / \beta)} \right)^2 
\label{G-phi-tau}
\end{eqnarray}
over the range
\begin{eqnarray}
|\tau| >> { 1 \over \Lambda} 
~~~~~{\rm and}~~~~~
|\beta - \tau| >> { 1 \over \Lambda} .
\label{tau-range}
\end{eqnarray}
We then have
\begin{eqnarray}
\tilde{\chi}^{(b)} (\tau) && =
- { {  g^2 } \over 8} e^{-\beta \lambda}
\int_0^{\beta-\tau}d\tau_1 \int_0^{\tau} d \tau_2
G^b_{\phi}(\tau_1 + \tau_2) \nonumber\\
&&
=
- {1 \over 4}
e^{-\beta \lambda}
(K_0g)^2 \ln \left(
{ { \sin (\pi \tau/\beta) }
\over 
\pi/\beta \Lambda }
\right) ,
\label{chi-tilde-b2}
\end{eqnarray}
where the factor $\Lambda$ appears due to the fact
that Eq. (\ref{G-phi-tau}) is valid only in the range
specified by Eq. (\ref{tau-range}).
Similarly,
\begin{eqnarray}
\tilde{\chi}^{(cd)} (\tau) && =
{ { 3 g^2 } \over 4} e^{-\beta \lambda}
\int_0^{\tau}d\tau_1 \int_0^{\tau-\tau_1} d \tau_2
G^b_{\phi}(\tau-\tau_1 - \tau_2) \nonumber\\
&& =
-{3 \over 4} e^{-\beta \lambda}
(K_0g)^2 \ln \left(
{ { \sin(\pi \tau /\beta)}
\over 
\pi/\beta 
 \Lambda }
\right) .
\label{chi-tilde-cd2}
\end{eqnarray}

\vskip 0.2 in

\section{Long-ranged one-dimensional spherical model with
a logarithmic-correction to its range}
\label{sec:appen-spherical}

Consider a spherical model, defined on a chain of $M$ sites, with a reduced
Hamiltonian
\begin{eqnarray}
{\cal H}_{sp} = -{ 1 \over 2} \sum_{ij} v_{ij} s_i s_j .
\label{H-sp}
\end{eqnarray}
Here, $s_i$ is a scalar variable constrained by the condition
\begin{eqnarray}
\sum_{i=1}^M s_i^2 = M ,
\label{constraint-sp}
\end{eqnarray}
with $M$ being taken to infinity at the end of the calculation.
The coupling $v_{ij}$ represents a long-ranged interaction whose
Fourier transform $v(p) \equiv \sum_{j} v_{ij} e^{-i p R_{ij}}$
is, for small $p$, either
\begin{eqnarray}
v(p) = v - v_1 |p|^{x} \quad \text{with $0<x<1$},
\label{vp-x}
\end{eqnarray}
or
\begin{eqnarray}
v(p) = v - { v_1 \over {\ln (1/|p|)}};
\label{vp-ln}
\end{eqnarray}
in either case, $v$ and $v_1$ are positive constants.

The long-range interaction corresponding to Eq.~(\ref{vp-x}) has the form
\begin{eqnarray}
v(r) \sim {1 \over {|r|^{1+x}}}.
\label{vr-x}
\end{eqnarray}
The solution to this problem is known.\cite{Joyce}
An unstable fixed point
separates a paramagnetic state
from an ordered state. The spin-spin correlation function
at the critical point behaves as
\begin{eqnarray}
\chi^{\text{cri}} (r) \sim {1 \over |r|^{1-x} } ,
\label{chi-r-vc}
\end{eqnarray}
while that on the paramagnetic side varies as
\begin{eqnarray}
\chi^{\text{para}} (r) \sim {1 \over |r|^{1+x} } .
\label{chi-r-v-lt-vc}
\end{eqnarray}
Eq.~(\ref{chi-r-v-lt-vc}) implies that on the paramagnetic side,
the exponent of the spin-spin correlation function saturates the bound
set by Griffiths' theorem.\cite{Griffiths}
The result is also consistent with those for the paramagnetic side of any 
generic one-dimensional classical spin models with power-law long-ranged
interactions.\cite{Fisher,Suzuki,Kosterlitz}

We now consider the problem with a long-range interaction
specified by Eq.~(\ref{vp-ln}). In real space, $v(r)$
has a $1/|r|$ dependence with a logarithmic correction.
We expect that, at the critical point, the spin-spin correlation
function should have a pure power-law decay without any logarithmic
correction. In the following, we show that this is indeed the case.
Standard manipulation\cite{Joyce,BerlinKac} leads to the following form
for the free-energy:
\begin{eqnarray}
F / M = \lambda_0 - { 1 \over 2} \int d p \,\ln \left(\lambda_0 -
v + { v_1 \over {\ln (1/|p|)}} \right),
\label{F-sp}
\end{eqnarray}
where the Lagrange multiplier, $\lambda_0$, 
introduced to enforce the constraint,
Eq.~(\ref{constraint-sp}), is determined by minimizing $F$:
\begin{eqnarray}
{1 \over 2} \int { d p \over
{ \lambda_0 -
v +  v_1 / \ln (1/|p|)}} = 1 .
\label{saddle-sp}
\end{eqnarray}
The critical point occurs when the solution is such that $\lambda_0 = v$,
corresponding to
\begin{eqnarray}
{1 \over 2} \int d p
{ {\ln (1/|p|)} \over
v_1^c} = 1 .
\label{saddle-sp-critical}
\end{eqnarray}
Note that the integral on the left hand side 
is infrared convergent, establishing
the existence of a phase transition.
The correlation function at the critical point is given by
\begin{eqnarray}
\chi^{\text{cri}} (p) \sim { 1 \over v_1^c} \ln{ 1 \over |p|} .
\label{chi-p-vc-ln}
\end{eqnarray}
In real space, this corresponds to
\begin{eqnarray}
\chi^{\text{cri}} (r) \sim { 1 \over |r|}
\label{chi-r-vc-ln}
\end{eqnarray}
without any logarithmic correction, which is what we set out to establish.


\begin{references}
\bibitem{note-dimensionality} The situation appears to be rather similar to
that in the insulating cuprates, where the fluctuations in the paramagnetic
phase are essentially two-dimensional (except for intra-unit-cell
couplings in bilayered cuprates) but a tiny $J_{\perp}$ (about
$10^{-5} J_{\parallel}$) produces three-dimensional
ordering due to a large in-plane correlation length.
\bibitem{E-loc-ordered}
The Kondo effect inside the ordered phase has been addressed in 
Ref.~\onlinecite{Zhu:03}.
Once $E_{\text{loc}}^*$ vanishes at the QCP it should remain
equal to zero
inside the antiferromagnetically ordered state.
The magnetic ordering removes some spectral weight from
the dynamical part of the Weiss field, replacing
it by a static Weiss field. Since a static field is more
efficient at destroying the Kondo effect than
the corresponding dynamical Weiss field,
the Kondo effect
should
remain inoperative inside the ordered phase.
\end{references}
\end{document}